%% file: 0-main.tex
\pdfoutput=1 
\documentclass{bmcart}

\usepackage{amsthm,amsmath}
\usepackage{amsfonts} 
\usepackage{graphicx} 
\usepackage{algpseudocode} 
\usepackage{booktabs} 
\usepackage{subcaption} 
\RequirePackage{natbib}
\RequirePackage{hyperref}
\usepackage[utf8]{inputenc} 

\startlocaldefs
\endlocaldefs

\newcommand{\modelname}{{PEM}}
\newcommand{\mytitle}{
Detecting Political Biases of Named Entities and Hashtags on Twitter
}

\newtheorem{definition}{Definition}
\newtheorem{algorithm}{Algorithm}

\begin{document}

\begin{frontmatter}

\begin{fmbox}
\dochead{Regular Article}


\title{\mytitle}


\author[
   addressref={aff1},                   
   email={patricia.xiao@cs.ucla.edu}   
]{\inits{ZX}\fnm{Zhiping} \snm{Xiao}}
\author[
   addressref={aff1},                   
   email={jeffrey7221@gmail.com}   
]{\inits{JZ}\fnm{Jeffrey} \snm{Zhu}}
\author[
   addressref={aff1},                   
   email={wangyining@g.ucla.edu}   
]{\inits{YW}\fnm{Yining} \snm{Wang}}
\author[
   addressref={aff2},                   
   email={peiz@usc.edu}   
]{\inits{PZ}\fnm{Pei} \snm{Zhou}}
\author[
   addressref={aff1},                   
   email={wenhong@g.ucla.edu}   
]{\inits{WHL}\fnm{Wen Hong} \snm{Lam}}
\author[
   addressref={aff3,aff4},                   
   email={mason@math.ucla.edu}   
]{\inits{MAP}\fnm{Mason A.} \snm{Porter}}
\author[
   addressref={aff1},                   
   email={yzsun@cs.ucla.edu}   
]{\inits{YS}\fnm{Yizhou} \snm{Sun}}


\address[id=aff1]{
  \orgname{Department of Computer Science, University of California, Los Angeles}, 
  \street{580 Portola Plaza},                     %
  \postcode{90095}                                
  \city{Los Angeles},                              
  \state{California},
  \cny{United States of America}                                    
}
\address[id=aff2]{%
  \orgname{Information Sciences Institute, University of Southern California},
  \street{Marina del Rey},
  \postcode{90292}
  \city{Los Angeles},
  \state{California},
  \cny{United States of America}
}
\address[id=aff3]{%
  \orgname{Department of Mathematics, University California, Los Angeles},
  \street{520 Portola Plaza},
  \postcode{90095}
  \state{California},
  \cny{United States of America}
}
\address[id=aff4]{%
  \orgname{Santa Fe Institute},
  \street{1399 Hyde Park Road},
  \postcode{87501}
  \city{Santa Fe},
  \state{New Mexico},
  \cny{United States of America}
}



\end{fmbox}


\begin{abstractbox}

\begin{abstract} 
Ideological divisions in the United States have become increasingly prominent in daily communication. Accordingly, there has been much research on political polarization, including many recent efforts that take a computational perspective. 
By detecting political biases in a corpus of text, one can attempt to describe and discern the polarity of that text.
Intuitively, the named entities (i.e., the nouns and the phrases that act as nouns) and hashtags in text often carry information about political views. 
For example, people who use the term ``pro-choice'' are likely to be liberal, whereas people who use the term ``pro-life'' are likely to be conservative.
In this paper, we seek to reveal political polarities in social-media text data and to quantify these polarities by explicitly assigning a polarity score to entities and hashtags.
 Although this idea is straightforward, it is difficult to perform such inference in a trustworthy quantitative way.
Key challenges include the small number of known labels, the continuous spectrum of political views, and the preservation of both a polarity score and a polarity-neutral semantic meaning in an embedding vector of words. 
To attempt to overcome these challenges, we propose the \textbf{P}olarity-aware \textbf{E}mbedding \textbf{M}ulti-task learning (\textbf{\modelname}) model. This model consists of (1) a self-supervised context-preservation task, (2) an attention-based tweet-level polarity-inference task, and (3) an adversarial learning task that promotes independence between an embedding's polarity dimension and its semantic dimensions.
Our experimental results demonstrate that our \textbf{\modelname} model can successfully learn polarity-aware embeddings that perform well at tweet-level and account-level classification tasks.
We examine a variety of applications --- including spatial and temporal distributions of polarities and a comparison between tweets from Twitter and posts from Parler --- and we thereby demonstrate the effectiveness of our \textbf{\modelname} model. We also discuss important limitations of our work and encourage caution when applying the \textbf{\modelname} model to real-world scenarios.
\end{abstract}


\begin{keyword}
\kwd{data sets}
\kwd{word embeddings}
\kwd{multi-task learning}
\kwd{adversarial training}
\end{keyword}


\end{abstractbox}
%

\end{frontmatter}



\input{1-introduction.tex}

\input{2-related.tex}

\input{3-definition.tex}
\input{4-model.tex}
\input{5-experiments.tex}

\input{6-conclusion.tex}

\input{7-acknowledgement.tex}

\bibliographystyle{bmc-mathphys} 
\bibliography{bmc_article}      

\end{document}

%% file: 1-introduction.tex
\section{Introduction}

In the United States, discourse has seemingly become very polarized politically and it often seems to be divided along ideological lines~\cite{levendusky2009partisan,webster2017ideological}. This ideological division has become increasingly prominent, and it influences daily communication.

\begin{figure}[h]
\centering
\includegraphics[width=0.9\columnwidth]{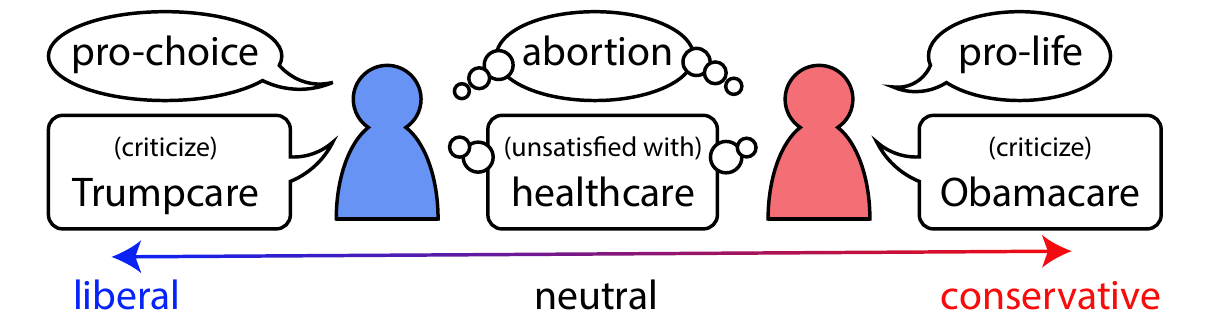}
\caption{Illustration of inferring political polarities from text.
}
\label{fig:inferring}
\end{figure}

The analysis of data from social media is important for studying human discourse~\cite{schober2016social,chao2022inference}. To study the polarization of social opinions in online communication, we attempt to detect polarity biases of entities and hashtags. There are a variety of ways to model political biases; see, e.g., VoteView~(see \url{https://voteview.com/})~\cite{boche2018new}. A space of political opinions can include axes for social views (e.g., ranging from ``conservative'' to ``progressive''), economic views (e.g., ranging from ``socialist'' to ``capitalist''), views on government involvement (e.g., ranging from ``libertarian'' to ``authoritarian''), and many others. The simplest model of a political spectrum, which we use in the present paper, is to consider a one-dimensional (1D) political space with views that range from ``liberal'' to ``conservative''.

By glancing at a corpus of text (such as a newspaper article or a tweet), humans can often readily recognize particular views in it without the need to analyze every word in the corpus.
Many items (including named entities and hashtags) in a corpus of text are helpful for inferring political views~\cite{gentzkow2010drives}, and
 people can quickly discern political views even in small corpora of text or in short speeches. 

On Twitter, political biases are often reflected in the entities and hashtags in tweets. The entities that we use are nouns and noun phrases (i.e., phrases that act as nouns), which we identify from text corpora by using existing natural-language-processing (NLP) tools.
 For instance, as we illustrate in Figure~\ref{fig:inferring}, if somebody uses the term ``pro-choice'' to describe abortion, they may have a liberal-leaning stance on a liberal--conservative axis of political views~\cite{rye2020pro}. By contrast, if somebody uses the term ``pro-life'', perhaps they have a conservative-leaning stance. We propose to automate this process in an interpretable way by detecting the political biases of entities and hashtags, inferring their attention weights in tweets, and then inferring the political polarities of tweets. 

The problem of inferring political polarities from text is somewhat reminiscent of ``fairness-representation'' problems~\cite{zhao2018learning,bose2019compositional}. This analogy is not perfect, and these problems have different objectives. We aim to reveal polarities, whereas fairness studies are typically interested in removing polarities. The notion of fairness entails that outputs are unaffected by personal characteristics such as gender, age, and place of birth. In recent studies, Zhao et al.~\cite{zhao2018learning} examined how to detect and split gender bias from word embeddings and Bose and Hamilton~\cite{bose2019compositional} developed models to hide personal information (such as gender and age) from the embeddings of nodes in graph neural networks (GNNs). Political bias can be more subtle and change faster than other types of biases. A key challenge is the labeling of political ideologies. Unlike the inference of gender bias, where it is typically reasonable to use discrete (and well-aligned) word pairs such as ``he''/``she'' and ``waiter''/``waitress'' as a form of ground truth, political polarity includes many ambiguities~\cite{tayal2014polarity}. Political ideology exists on a continuous spectrum, with unclear extremes, so it is very hard to determine either ground-truth polarity scores or well-aligned word pairs (e.g., ``he'' versus ``she'' is aligned with ``waiter'' versus ``waitress'')~\cite{pla2014political}. 

To infer polarities,
we seek to learn an embedding that can help reveal both the semantic meaning and the political biases of entities and hashtags. We propose a model, which we call the \textbf{P}olarity-Aware \textbf{E}mbedding \textbf{M}ulti-task learning (\textbf{\modelname}) model, that involves three tasks: (1) preservation of the context of words; (2) preservation of corpus-level polarity information; and (3) an adversarial task to try to ensure that the semantic and polarity components of an embedding are as independent of each other as possible. 

Our paper makes the following contributions:
\begin{enumerate}
    \item[(1)] We raise the important and practical problem of studying political bias in a corpus of text, and we assemble a data set from Twitter to study this problem. 
    Our code, the data sets of the politicians, and the embedding results of our models are available at \url{https://bitbucket.org/PatriciaXiao/pem/src/master/}.
    \item[(2)] We propose the \textbf{\modelname} model to simultaneously capture both semantic and political-polarity meanings. 
    \item[(3)] Our \textbf{\modelname} model does not rely on word pairs to determine political polarities. Consequently, it is flexible enough to adapt to other types of biases and to use in other context-preservation strategies.
    \item[(4)] Our data, source code, and embedding results are helpful for tasks such as revealing potential political polarities in a text corpus. 
\end{enumerate}


%% file: 2-related.tex
\section{Related Work and Preliminary Discussions}\label{sec::related}


\subsection{Political-Polarity Detection}

There are a variety of ways to formally define the notion of political polarity~\cite{boche2018new}. We consider a 1D axis of political views that range from ``liberal'' to ``conservative''. In the United States, members of the Democratic party tend to be liberal and members of the Republican party tend to be conservative~\cite{levendusky2009partisan,lieberman2017trumpism}. This prior knowledge is helpful for acquiring high-quality labeled data~\cite{10.1145/3394486.3403275}, but such data are restricted in both amount and granularity.

The detection of political polarity has been a topic of considerable interest for many years~\cite{pierce1988two,maynard2011automatic}. 
Additionally, for more than a decade, social-media platforms like Twitter have simultaneously been an important source of political opinion data and have themselves impacted political opinions in various ways~\cite{barbera2014social,bail2018exposure}. Some researchers have attempted to infer the political views of Twitter accounts from network relationships (such as following relationships)~\cite{gu2016ideology,tien2020online,10.1145/3394486.3403275}. Other researchers have attempted to infer polarity from tweet text~\cite{iyyer2014political,lai2019stance}.

We seek to infer the political polarities of entities and hashtags in tweets. Gordon et al.~\cite{gordon2020studying} illustrated recently that word embeddings can capture information about political polarity, but their 
approach does not separate polarity scores from embeddings and thus cannot explicitly tell which words are biased. Most prior research has focused on tweet-level or account-level polarities~\cite{vergeer2015twitter,jungherr2016twitter} or on case studies of specific ``representative'' hashtags~\cite{powell2022hashtags}.
By contrast, our \textbf{\modelname} model focuses on biases at a finer granularity (specifically, entities and hashtags). 

\subsection{Neural Word Embeddings}

We use the term \textit{neural word embeddings} to describe approaches to represent tokens (e.g., words) using vectors to make them understandable by neural networks~\cite{bengio2003neural,levy2014neural,li2015word}. Words can have very different meanings under different tokenizations. In our paper, we tokenize text into entities (including nouns and noun phrases), hashtags, emoji, Twitter handles, and other words (including verbs, adjectives, and so on). One way to obtain a neural word embedding is the {\sc Skip-gram} version of {\sc word2vec} approaches~\cite{mikolov2013distributed}, which are based on the assumption that similar words have similar local textual contexts. 
Another approach, which is called {\sc GloVe}~\cite{pennington2014glove}, relies on a global co-occurrence matrix of words.
Other methods, such as transformers~\cite{vaswani2017attention,devlin2018bert}, generate contextualized embeddings (in which a word can have different embeddings in different contexts). These models encode words, which initially take the form of a sequence of characters, into a vector space. Therefore, these models are also often called ``encoders''.

In contrast to all of the above studies, our \textbf{\modelname} model learns an embedding that captures both the semantic meanings and the political polarities of words. Our framework is not limited to any specific embedding strategy. If desired, one can replace the embedding part (namely, Task \#1) of our \textbf{\modelname} model by other encoders.


\subsection{Fairness of Representations}

Many researchers have observed that word embeddings often include unwanted biases~\cite{mehrabi2019survey}. In studies of fairness, a model is considered to be ``fair'' if its outputs are unaffected by personal characteristics, such as gender and age; it is ``biased'' (i.e., ``unfair'') if such features influence the outputs. Models often inherit biases from training data sets, and they can exacerbate such biases~\cite{o2016weapons}.
Researchers have undertaken efforts to reveal biases and mitigate them~\cite{bose2019compositional}. For example, Zhao et al. revealed gender-bias problems using their {\sc WinoBias} model~\cite{zhao2018gender} and attempted to generate gender-neutral representations using their {\sc GN-GloVe} model~\cite{zhao2018learning}.

Such representation-learning algorithms motivate us to separate politically-biased and politically-neutral components in embeddings (see \cite{zhao2018learning}) and to use an adversarial training framework to enhance the quality of the captured polarities (see \cite{bose2019compositional}). However, our work has a different focus than \cite{zhao2018learning} and \cite{bose2019compositional}. These works were concerned with reducing biases, whereas we seek to reveal differences between polarized groups.


\subsection{Sentiment Analysis}

Sentiment analysis aims to determine the attitude (negative, positive, or neutral) of a corpus of text~\cite{medhat2014sentiment,astya2017sentiment}.
The use of neural word embeddings is common in statistical approaches to sentiment analysis~\cite{yu2017refining,fu2018learning}. Some of these approaches account for the importance levels of entities~\cite{batra2010entity,song2017semi}. 

In many applications, sentiment analysis has relied on much richer labeled data sets than those that are available in political contexts ~\cite{tang2014learning,astya2017sentiment}, where it is rare to find high-quality anchor words (such as good, bad, like, and dislike)~\cite{yu2017refining}. In our paper, we seek to reveal polarities from textual data. Polarity is different from sentiment. For example, most entities have neutral sentiments, but these same entities can still have biased polarities.


\subsection{Recognition of Named Entities}

We focus on learning polarity scores for named entities (specifically, nouns and noun phrases) and hashtags. The terminology ``named entity'', which comes from NLP, refers to a noun or a noun phrase that is associated with an entity. For example, the \textit{United States Congress} is a named entity.  We use a named-entity recognition (NER) tool ~\cite{nadeau2007survey,li2020survey} to identify the entities in our training corpus. In an NER information-extraction task, one seeks to discern and classify entities in a text corpus into predefined categories, such as person names, organizations, and locations. We use the popular tools {\sc TagMe}~\cite{ferragina2010tagme} and {\sc AutoPhrase}~\cite{shang2018automated} for our tasks.

%% file: 3-definition.tex
\section{Problem Definition}\label{sec::def}

 We use ``tokens'' to denote the smallest word units that we obtain through tokenization of tweets. We tokenize entities, hashtags, emoji, mentioned accounts, and other words. We represent each tweet as a sequence of such tokens. We study the problem of detecting the political biases of entities and hashtags in tweets. To do this, we seek to learn (1) semantic embeddings for each token and (2) the political polarities of each entity and hashtag. We then obtain tweet-level polarity scores by calculating a weighted average of token-level polarity scores.

\begin{definition}{\textbf{(Two-Component Polarity-Aware Embeddings)}}
We design a two-component polarity-aware embedding $\mathbf{z} \in \mathbb{R}^{d_1+d_2}$ of each token $\mathbf{w}$. Because we seek to learn 1D polarity scores, we set $d_2 = 1$. We decompose $\mathbf{z}$ as follows:
\begin{displaymath}
    \mathbf{z} = [\mathbf{z}^{(s)}\,,\, \mathbf{z}^{(p)}]\,,\quad \mathbf{z}^{(s)} \in \mathbb{R}^{d_1}\,,\, \mathbf{z}^{(p)} \in \mathbb{R}^{d_2}\,.
\end{displaymath}
The two components of the embedding $\mathbf{z}$ are
\begin{itemize}
    \item[(1)] the \textbf{polarity-neutral} semantic component $\mathbf{z}^{(s)}$ and
    \item[(2)] the \textbf{polarity-aware} political-polarity component $\mathbf{z}^{(p)}$.
\end{itemize}
\end{definition}

By forcing $\mathbf{z}^{(s)}$ to be polarity-neutral, we seek to enhance the quality of the political polarities that we capture in $\mathbf{z}^{(p)}$. We set $d_1 = d$ and $d_2 = 1$, and we use $f(\mathbf{z}^{(p)}) = z_{d+1}$ as the ``polarity score'' of a token. When determining tweet-level polarities, we ignore $\mathbf{z}^{(p)}$ for tokens that are neither entities nor hashtags. We expect that $z_{d+1} < 0$ when a word is liberal-leaning and that $z_{d+1} > 0$ when a word is conservative-leaning. The absolute value $|z_{d+1}|$ indicates the magnitude of a political leaning. Using our approach, we are able to infer the political polarity of a token in $\mathcal{O}(1)$ time. 
We are interested in the polarity scores of tokens that are either entities or hashtags. It is very common to use a 1D polarity score~\cite{boche2018new}, so we do so in the present paper. However, it is straightforward to extend our \textbf{\modelname} model to incorporate more polarity dimensions.

%% file: 4-model.tex
\section{Methodology}\label{sec::methodology}

\subsection{General Design}

\begin{figure}[h]
\centering
\includegraphics[width=0.9\columnwidth]{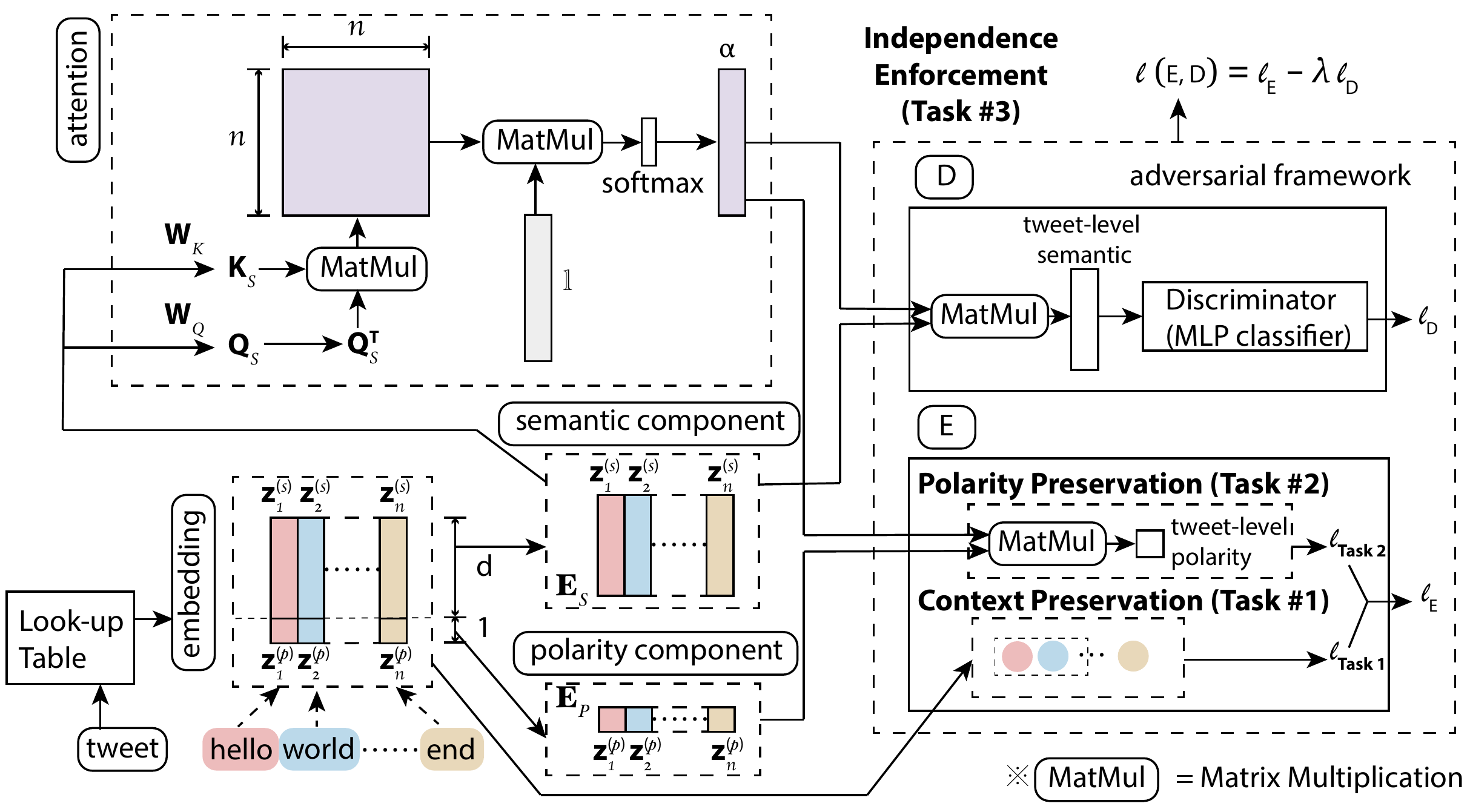}
\caption{Schematic illustration of our \textbf{\modelname} model. In this illustration, we consider a tweet with $n$ tokens.
}
\label{fig:illustration}
\end{figure}

To generate our proposed embeddings, we infer semantic meanings, infer political polarities, and use $\mathbf{z}^{(p)}$ to capture as much political polarity as possible. 

We show a schematic illustration of our model in Figure \ref{fig:illustration}. To capture the meanings of tokens, we learn embeddings from the context of text. We thus propose Task \#1 to help preserve contextual information. To infer political polarities from tokens, we propose Task \#2, in which we use a weighted average of the entities' and hashtags' polarity component $\mathbf{z}^{(p)}$ to calculate a polarity score of each tweet. To further enhance the quality of the polarity component, we propose Task \#3, in which we use an adversarial framework to ensure that the two components, $\mathbf{z}^{(s)}$ and $\mathbf{z}^{(p)}$, are as independent as possible.


\subsection{Task \#1: Context Preservation}

We want our token-level embeddings to preserve contextual information, which has both semantic information and polarity information.
A simple approach is to use {\sc Skip-Gram}~\cite{mikolov2013distributed}. Given a document with tokens $w_1, w_2, \ldots, w_n$, we seek to maximize the mean log probability to observe tokens in a local context. Specifically, we maximize
\begin{equation}\label{eq:original_skipgram}
    \frac{1}{n} \sum_{t = 1}^{n} \sum_{j \in \{-c, \ldots, c\},\, j \neq 0} \ln p(w_{t+j}|w_t)\,,
\end{equation}
where $c$ indicates the size of a sliding window and 
\begin{equation}\label{index}
    p(w_{t+j} | w_t) = \frac{\exp(\mathbf{z}_{t}^T \mathbf{z}_{t+j}')}{\sum_{i = 1}^{|\mathbf{W}|} \exp(\mathbf{z}_{t}^T \mathbf{z}_{i}')}\,,
\end{equation}
where $w_i$ is the $i$th token in the document, the set $\mathbf{W}$ is the vocabulary set of all tokens,
$\mathbf{z}_i$ is the target embedding of token $w_i$, and $\mathbf{z}_{i}'$ is the context embedding.
When the index $t+j \not \in \{1, \ldots, n\}$, we ignore it in \eqref{index}.
In Task\# 1, we need both
$\mathbf{z}_i$ and $\mathbf{z}_{i}'$ to be able to distinguish between the target and context roles of the same token~\cite{mikolov2013distributed}. In Task \#2 (see Section \ref{task2}) and Task \#3 (see Section \ref{task3}), we use only the context embedding $\mathbf{z}_{i}'$.

The loss function $\ell_{\text{Task 1}}$ for Task \#1 is the negative-sampling objective function
{\small
\begin{equation}\label{eq:loss_task1}
    \ell_{\text{Task 1}} = -\frac{1}{k+1}\left( \ln \left(\sigma (\mathbf{z}_{t}^T \mathbf{z}_{t+j}')\right) + \sum_{i = 1}^k \mathbb{E}_{w_i \sim P_{\mathrm{noise}}(w)}\left[ \ln\left(\sigma(-\mathbf{z}_{t}^T \mathbf{z}_{i}')\right) \right]\right) \,,
\end{equation}}\!\!
where $k$ is the number of negative samples (i.e., token pairs that consist of a target token and a token from a noise distribution) per positive sample (i.e., token pairs that occur in the same sliding window), the sigmoid function $\sigma$ is $\sigma(x) = \frac{1}{1+\exp(-x)}$, and $P_{\mathrm{noise}}(\cdot)$ is a noise distribution. We obtain negative samples of word pairs from the noise distribution~\cite{mikolov2013distributed}, whose name comes from the idea of noise-contrastive estimation (NCE)~\cite{pmlr-v9-gutmann10a}. A good model should distinguish between data and noise. 
We use the same noise distribution as in {\sc Skip-Gram}~\cite{mikolov2013distributed}:
\begin{equation}
    P_{\mathrm{noise}}(w) = \left( \frac{U(w)}{\sum_{i \in \mathbf{W}} U(i)} \right)^{3/4}\,,
\end{equation}
where $U(w)$ denotes the number of appearances of a token $w$ in the training corpus.
Minimizing $\ell_{\text{Task 1}}$ 
approximates the maximization of 
the mean log probability \eqref{eq:original_skipgram}.

In practice, when discussing political affairs, they are usually described by multiple words, namely, phrases. We use {\sc AutoPhrase}~\cite{shang2018automated} to detect phrases in our data sets, and treat them as tokens as well. 

We refer to Task \#1 as our \textbf{Baseline} \textbf{\modelname} model, and we call it the ``{\sc Skip-Gram} model'' when we use it on its own. We use the same hyperparameter settings as in the default settings in the original {\sc Skip-Gram} model~\cite{mikolov2013distributed}.


\subsection{Task \#2: Polarity Preservation} \label{tasktwo}

In Task \#2, our goal is for the polarity component of our embeddings to capture reasonable polarity information. The finest granularity of the polarity labels that we can automatically and reliably obtain in large enough numbers are at the level of social-media accounts. We assume that every politician has consistent political views during our observation time (the years 2019 and 2020), and we assign polarity labels to their tweets based on their self-identified party affiliations. We thereby use account-level labels to guide the polarity-score learning of entities and hashtags.

A simple approach is to use the mean polarity score of all entities to estimate the polarity score of a text corpus. However, this approach does not consider the heterogeneous importance levels of entities. When considering political tendencies, some entities (e.g., ``pro-choice'') are more informative than others (e.g., ``plan''). Therefore, we calculate a weighted average of entity polarities in each tweet, with weights that come from attention.

Suppose that we are given a sentence with $n$ tokens (i.e., words, phrases, hashtags, mentions, emoji, and so on) that are embedded as $\mathbf{z}_1, \mathbf{z}_2, \ldots, \mathbf{z}_n$, where $m$ of the $n$ tokens are entities or hashtags. The set of indices of the $m$ tokens is $\mathbf{I} = \{i_1, \ldots, i_m\}$ (with $m \leq n$). The polarity dimensions of the embeddings are 
\begin{displaymath}
    \mathbf{E}_P = [\mathbf{z}_{i_1}^{(p)}\,;\, \mathbf{z}_{i_2}^{(p)}\,;\,  \cdots \,;\, \mathbf{z}_{i_m}^{(p)}] \in \mathbb{R}^{m \times 1}\,.
\end{displaymath}

We use a standard self-attention mechanism~\cite{hu2019introductory}, which proceeds as follows. We represent keys, values, and queries in a vector space. Each key has a corresponding value. Upon receiving a query, we evaluate similarities between the queries and the keys. We then estimate the value of a query as a weighted average of the values that correspond to the keys~\cite{vaswani2017attention}.

We vertically concatenate the sequence of the semantic (i.e., polarity-neutral) components of the entities' and hashtags' embeddings and write
\begin{displaymath}
    \mathbf{E}_S = [\mathbf{z}_{i_1}^{(s)}\,;\,  \mathbf{z}_{i_2}^{(s)}\,;\,  \cdots \,;\, \mathbf{z}_{i_m}^{(s)}] \in \mathbb{R}^{m \times d}\,,
\end{displaymath}
where the key $\mathbf{K}$ and the query $\mathbf{Q}$ are different linear transformations of $\mathbf{E}_S$. 
That is,
\begin{displaymath}
    \mathbf{K} = \mathrm{stopgrad}\left( \mathbf{E}_{S}  \right) \mathbf{W}_{K}\,, \quad \mathbf{Q} = \mathrm{stopgrad}\left( \mathbf{E}_{S} \right) \mathbf{W}_{Q}\,,
\end{displaymath}
where $\mathrm{stopgrad}$ is a stop gradient (so $\mathbf{E}_{S}$ is not updated by back-propagation of the attention component)
and $\mathbf{W}_{K}, \mathbf{W}_{Q} \in \mathbb{R}^{d \times d}$ are weight matrices. We calculate the attention vector $\boldsymbol{\alpha} \in \mathbb{R}^{m \times 1}$, which includes an attention score for each entity in a tweet, using the standard softmax function:
\begin{equation}
    \boldsymbol{\alpha} = \mathrm{Att}(\mathbf{Q},\mathbf{K}) = \mathrm{softmax}\left(\left( \frac{\mathbf{Q} \mathbf{K}^T}{\sqrt{m}} \right) \cdot\mathbf{1}_{m\times1} \right)\,,
\end{equation}
where the $i$th component of the softmax function is
\begin{equation*}
    \mathrm{softmax}(x_i) = \frac{e^{x_i}}{\sum_{k = 1}^m e^{x_k}}
\end{equation*}
and $\mathbf{1}_{m\times1}$ is a vector of 
$1$ entries.

Each tweet's polarity score $\tilde{\mathbf{z}}^{(p)}$ is then
\begin{equation}
    \tilde{\mathbf{z}}^{(p)} = \boldsymbol{\alpha}^T \mathbf{E}_P \in \mathbb{R}^{1 \times 1}\,.
\end{equation}
Suppose that there are $N$ tweets in total and that tweet $j$ has the associated label $l_j \in\{-1,1\}$, where $-1$ signifies that the tweet is by a politician from the Democratic party and $1$ signifies that the tweet is by a politician from the Republican party. (We only consider politicians with a party affiliation.)
We infer polarity scores $\{ \tilde{\mathbf{z}}^{(p)}_1, \tilde{\mathbf{z}}^{(p)}_2, \ldots, \tilde{\mathbf{z}}^{(p)}_N \}$ for each tweet and then use a hinge loss with the margin parameter $\gamma > 0$ as our objective function.
Specifically, we set $\gamma = 1$ and write the loss for Task \#2 as
\begin{equation}
    \ell_{\text{Task 2}} = \frac{1}{N} \sum_{j = 1}^N \left( \max\left\{ 0, \gamma - l_j \tilde{\mathbf{z}}^{(p)}_j \right\} \right)\,.
\end{equation}

When we use Task \#1 and Task \#2, we say that we are using our \textbf{Polarized} \textbf{\modelname} model.


\subsection{Task \#3: Independence Enforcement} \label{task3}

In Task \#3, we encourage the semantic component $\mathbf{z}^{(s)}$ to be polarity-neutral, and we thereby force the political-polarity component $\mathbf{z}^{(p)}$ to capture polarity more accurately. We use an adversarial framework to achieve this goal. We alternately train two competing objectives: (1) learn a high-quality embedding $\mathbf{z}$ that preserves both context and polarity; and (2) learn a semantic embedding $\mathbf{z}^{(s)}$ that is not able to infer a tweet's polarity. Let $E$ denote the first objective, which combines Task \#1 and Task \#2 and controls the quality of our embedding. The loss function $\ell_{E}$ of the first objective is
\begin{equation}
    \ell_{E} = \ell_{\text{Task 1}} + \ell_{\text{Task  2}}\,.
\end{equation}
Let $D$ denote the second objective, which is a discriminator that attempts to use a semantic embedding for polarity classification. We start training by running the objective $E$ because our discriminator makes sense only if our embedding is meaningful.

We apply the attention mechanism that we used in Task \#2 (for aggregate token-level semantic embeddings) to a tweet-level semantic embedding. We use the weighted average $\tilde{\mathbf{z}}^{(s)} = \boldsymbol{\alpha}^T  \mathbf{E}_S \in \mathbb{R}^d$ of the semantic dimensions of a tweet's tokens as our tweet-level semantic embedding. The $\mathbf{W}_K$ and $\mathbf{W}_Q$ functions in Task \#3 are different than
those in Task \#2. We use the discriminator $D$ to discern political-party labels from $\tilde{\mathbf{z}}^{(s)}$. The discriminator is a standard two-layer multilayer perceptron (MLP) classifier that infers a class label $0$ for liberal-leaning tokens and a class label $1$ for conservative-leaning tokens. Between these two layers, we set the number of elements in the output of each hidden layer to $d_{\mathrm{MLP}} = 100$. We use a binary cross-entropy loss $\ell_D$.
The ground-truth labels of the tweets are $\mathbf{Y} = \{ y_1, \ldots, y_N \} \in \{0, 1\}^N$ and the inferred polarity scores are $\hat{\mathbf{Y}} = \{\hat{y}_1, \ldots, \hat{y}_N\}$. The output label of tweet $i$ is
\begin{equation}
    \hat{y}_i = D(\tilde{z}^{(s)}) = \sigma\left(\mathrm{MLP}(\tilde{z}^{(s)})\right) \in [0, 1]\,,
\end{equation}
where $\sigma$ is the sigmoid function.
The discriminator loss is the binary cross entropy
\begin{equation}\label{eq:l_D}
    \ell_D = -\frac{1}{N}\sum_{i=1}^N\left(y_i \ln(\hat{y}_i) + (1 - y_i) \ln (1 - \hat{y}_i)\right)\,.
\end{equation}

The encoder $E$ seeks to make $\ell_D$ large enough so that $\mathbf{z}^{(s)}$ tends to ignore political polarity. The discriminator $D$ seeks to make $\ell_D$ small enough to be a stronger discriminator.
To balance these goals, we use an adversarial framework. The training objective for all tasks together is
\begin{equation}
    \begin{aligned}
\ell_{\text{Task 3}} = & \min_E \max_D \left(\ell(E, D)\right)
= \min_E\max_D \left(\ell_{E} - \lambda \ell_D \right)\,.
    \end{aligned}
\end{equation}

We always train Task \#3 together with Tasks \#1 and \#2.
When we train all three tasks together, it is referred as the \textbf{Complete} \textbf{\modelname} model.


\subsection{Joint Training}

In Algorithm~\ref{alg:ourmodel}, we present our adversarial framework for our \textbf{Complete} \textbf{\modelname} model. An adversarial framework trains two neural networks together so that they counteract each other~\cite{goodfellow2014generative,chen2016infogan}.
The quantity $\theta_E$ denotes all of the parameters in Tasks \#1 and \#2, including all of the embedding weights $\mathbf{Z}$, the attention weights, and so on. The quantity $\theta_D$, which we use only in Task \#3, denotes the set of discriminator parameters. Each batch that we input into our \textbf{\modelname} model has data from $16$ tweets.

We learn all parameters in $\theta_E$ and $\theta_D$ during training, but we need to determine the hyperparameter $\lambda$. In our experiments, we examined $\lambda = 0.01$, $\lambda = 0.1$, $\lambda = 1$, and $\lambda = 10$.
Of these values, our \textbf{Complete \textbf{\modelname}} model performs the best for $\lambda = 0.1$, so we use $\lambda = 0.1$. 
When applying the \textbf{\modelname} model to another data set, one should carefully select a suitable value of $\lambda$.

\begin{algorithm}
\begin{algorithmic}
{\textbf{Complete} \modelname: Learning algorithm}
\Procedure{LearnEmbedding}{$\mathrm{Iter}$} 
\State $\mathbf{Z}\gets$ initialize the embeddings
\State Initialize the parameter $\lambda > 0$
\For{$i$ = $1, \ldots, \mathrm{Iter}$}
\While{not converged} \Comment{ train $\theta_E$, fix $\theta_D$}
\State sample from tweets
\State $\ell_E \gets \ell_{\text{Task 1}} + \ell_{\text{Task 2}}$
\State $\ell(E, D) \gets \ell_E - \lambda \ell_D$
\State update $\theta_E$ to minimize $\ell(E, D)$ 
\EndWhile\label{learn\modelname}
\While{not converged} \Comment{ train $\theta_D$, fix $\theta_E$}
\State sample from tweets
\State $\ell_D \gets $ Discriminator loss
\State update $\theta_D$ to minimize $\ell_D$
\EndWhile\label{learnD}
\EndFor
\State \textbf{return} $\mathbf{Z}$\Comment{
the learned embedding 
}
\EndProcedure
\end{algorithmic}\label{alg:ourmodel}
\end{algorithm}

In each phase (i.e., either training $\theta_D$ or training $\theta_E$), we stop training right after we first observe a drop in the $F_1$ score (which is is the harmonic mean of precision and recall) in the validation set. (Such a performance drop can be an indication of overfitting~\cite{cawley2010over}.) We then use the parameter values from just before the performance drop and proceed to the next phase.

%% file: 5-experiments.tex
\section{Experiments}\label{sec::exp}

\subsection{Data Sets}\label{subsec::dataset}

We start by collecting a list of Twitter accounts, including 585 accounts of legislators in the $115^{\mathrm{th}}$ and $116^{\mathrm{th}}$ Congresses,\footnote{See \url{https://www.congress.gov/members}.} the accounts of $8$ well-known news outlets (see Table~\ref{tab:news_agencies}), and the accounts of President Barack Obama, President Donald Trump, and their Cabinet members at the time (3 March 2019) that we first collected the data. Our data set consists of (1) the most recent 3,200 tweets of each account that we collected on 3 March 2019 and (2) the tweets of these
accounts that were posted between 1 January 2020 and 25 November 2020.

We select the news outlets from those with the most voters (i.e., participants who label the political polarity of news outlets on the AllSides Media Bias Ratings (see \url{https://www.allsides.com/media-bias/media-bias-ratings}). Previous studies have inferred the political polarities of news outlets from their content~\cite{chao2022inference,gruppi2022scilander}, and we seek to examine whether or not our model can also reveal political polarities.
The available political labels in the AllSides Media Bias Ratings are ``liberal'', ``somewhat liberal'', ``neutral'', ``somewhat conservative'', and ``conservative''.
We use the three liberal news outlets with the most votes, the three conservative news outlets with the most votes, and the neutral news outlet with the most votes.
We checked manually that the polarities of the Twitter accounts of these news outlets are consistent with the labels that we obtained from the AllSides Media Bias Ratings. When a news outlet has multiple Twitter accounts (e.g., {\tt @cnn} and {\tt @cnnpolitics}), we use the account with the most followers in early February 2020. On 10 February 2020, we finished collecting and sorting the media data.

We split the politicians' tweets (of which there are more than 1,000,000 in total) into training, validation, and testing sets in the ratio 8:1:1. 
We also use the tweets of the news outlets and those of the unobserved accounts as testing sets.

We also test our embedding on three existing data sets: the {\sc Election2020} data set~\cite{chen2020election2020}, which has 965,620,919 tweets that were collected hourly between March 2020 and December 2020; a {\sc Parler} data set from 6 Jan 2021 that has 1,384,579 posts;\footnote{This data set is available at the repository \url{https://gist.github.com/wfellis/94e5695eb514bd3ad372d6bc56d6c3c8}.} and the TIMME data set~\cite{10.1145/3394486.3403275}, which includes 2,975 Twitter accounts with location information and self-identified political-polarity labels (either Democratic or Republican). These Twitter accounts are not run by politicians and are never in a training data set. We thus refer to them as ``unobserved accounts''. We have access to the most recent 3,200 tweets in each Twitter account's timeline; we keep the tweets that they posted in 2020.


\begin{table}[h]
    \centering
    \caption{The selected news outlets and their political polarities. The label ``L'' denotes a liberal-leaning outlet, ``C'' denotes a conservative-leaning outlet, and ``N'' denotes a neutral outlet. These labels come from the AllSides Media Bias Ratings (see \url{https://www.allsides.com/media-bias/media-bias-ratings}).
    }
    {
    \begin{tabular}{r|r|c}
        \toprule
        Twitter Account & News Outlet & Polarity \\ \midrule
        {\tt @nytimes} & \emph{The New York Times} & L \\
        {\tt @guardiannews} & \emph{Guardian News} & L \\
        {\tt @cnn} & \emph{CNN} & L \\
        {\tt @csmonitor} & \emph{The Christian Science Monitor} & N  \\
        {\tt @amspectator} & \emph{The American Spectator} & C \\
        {\tt @foxnewsopinion} & \emph{Fox News Opinion} & C \\
        {\tt @nro} & \emph{National Review} & C \\ \bottomrule
    \end{tabular}
    }
    \label{tab:news_agencies}
\end{table}


\subsection{Entity Identification}

We use the union of the set of entities from three main sources to identify potential entities while training.

To detect nouns, we consider all nouns and proper nouns from parts-of-speech (POS) tagging\footnote{See \url{https://www.nltk.org/api/nltk.tag.html}.} to be reasonable entities.

To detect phrases that act as nouns, we use {\sc AutoPhrase} (version 1.7)~\cite{shang2018automated} to learn a set of phrases from all politicians' tweets in our data.
We then use this set of phrases when tokenizing all employed data sets. {\sc AutoPhrase} assigns a score in the interval $[0, 1]$ to each potential phrase, where a higher score indicates a greater likelihood 
to be a reasonable phrase.
After looking at the results, we manually choose a threshold of $0.8$, and we deem all multi-word noun phrases whose scores are at least this threshold to be of sufficiently high quality. 

To detect special terms that represent entities that may not yet be part of standard English, we apply {\sc TagMe} (version 0.1.3)~\cite{ferragina2010tagme} to our training set to include named entities that we are able to link to a Wikipedia page.


\subsection{Results}

\subsubsection{Polarity Component}\label{subsec:token}

We compute token-level polarity scores by examining the polarity component $\mathbf{z}^{(p)}$ of each embedding. We transform all tokens except mentions into lower-case versions. We do this because Twitter handles (i.e., user names) are case-sensitive, but upper-case and lower-case letters have the same meaning (and thus can be used as alternatives to each other) for other entities (including hashtags).

According to our results, of the entities and hashtags that politicians used in our data (which we collected in 2019 and 2020), the ones with the strongest liberal polarities are \textbf{\#trumpcare}, \textbf{\#actonclimate}, \textbf{\#forthepeople}, \textbf{\#getcovered}, and \textbf{\#goptaxscam}. The entities and hashtags with the strongest conservative polarities are \textbf{\#va10}, \textbf{\#utpol}, \textbf{\#ia03}, \textbf{\#tcot}, and \textbf{\#wa04}.

Our results illustrate that hashtags that refer to electoral districts can be strongly conservative-leaning. Politicians with different political leanings may use hashtags in different ways, and examining a hashtag that is associated with an electoral district is a good way to illustrate this. Additionally, conservative politicians may use a particular non-germane 
hashtag for certain content more often than liberal politicians. For example, some tweets that used \textbf{\#va10} contributed to a discussion of a \#VA10 forum that was hosted by the Republican party in Fauquier County ({\tt @fauquiergop}).

\begin{figure}[h]
\centering
\includegraphics[width=1.0\columnwidth]{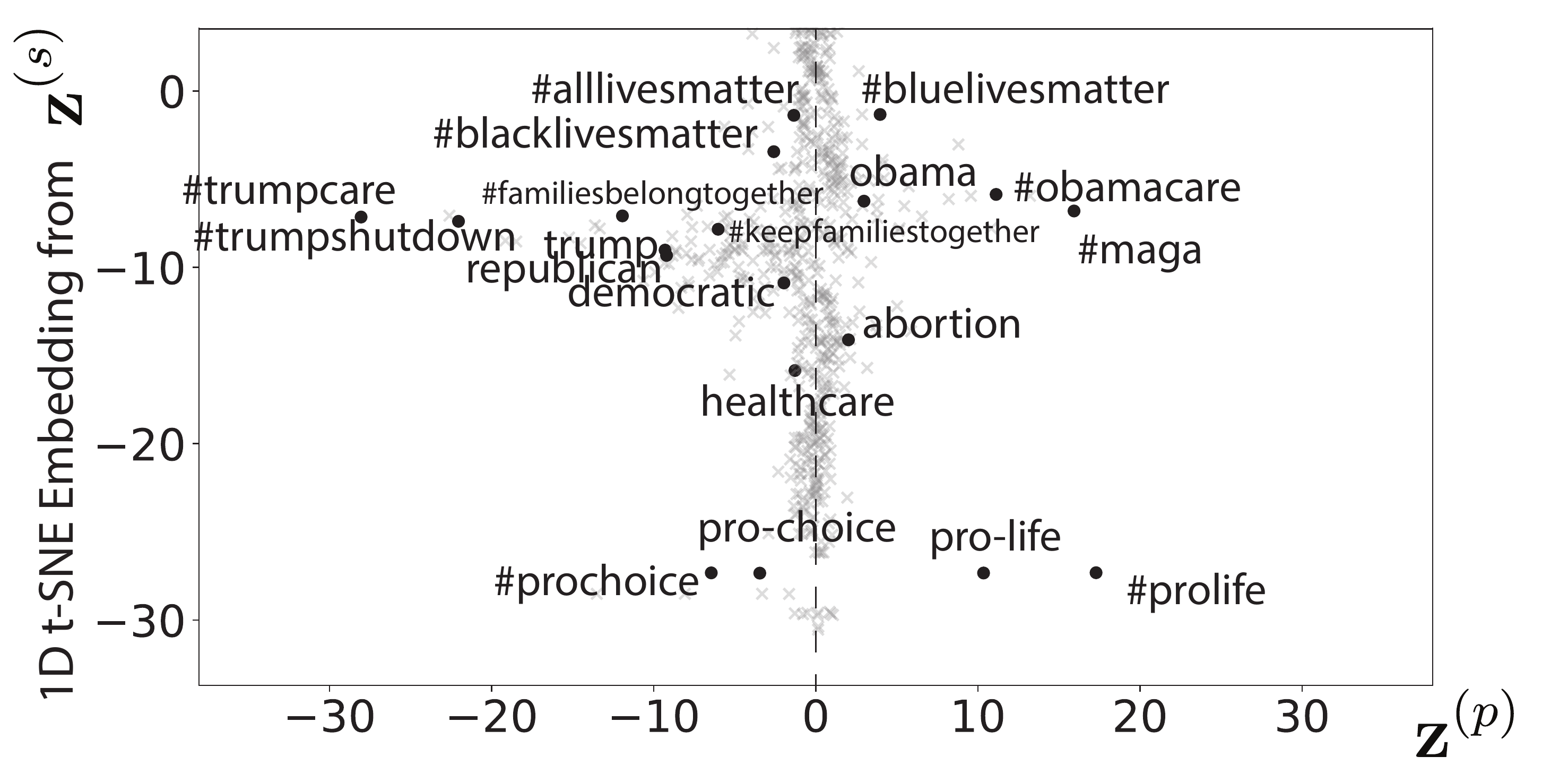}
\vspace{-.7cm}
\caption{Visualization of the political polarities in our embedding results. The horizontal axis gives the values of the polarity score $\mathbf{z}^{(p)}$, and the vertical axis is a 1D t-SNE value (which we use to facilitate visualization) that we calculate from the semantic embedding $\mathbf{z}^{(s)}$.
}
\label{fig:vis_polarity_dim_unnormalized}
\end{figure}

In Figure~\ref{fig:vis_polarity_dim_unnormalized}, we show our embedding results for the 1,000 most-frequent entities and hashtags and for a few highlighted ones that we select manually. To facilitate visualization, the vertical axis is a 1D t-distributed stochastic neighbor embedding (t-SNE) values~\cite{maaten2008visualizing}. In theory, words with particularly close semantic meanings are near each other along this axis. In our embedding results, hashtags are more likely than other tokens to capture a clear political polarity.

Some of our observations are unsurprising. For example, terms that are related to ``\textbf{pro-life}'' are typically conservative-leaning, whereas terms that are related to ``\textbf{pro-choice}'' are typically liberal-leaning.

Other observations are more nuanced. For example, liberal-leaning Twitter accounts sometimes use text that one is likely to associate more with conservative-leaning views, and vice versa. The embeddings of ``\textbf{trump}'' and ``\textbf{obama}'' give one pair of examples, and the hashtags \textbf{\#trumpcare} and \textbf{\#obamacare} give another. Hashtags without semantic context can also appear in tweets. Another interesting observation is that \textbf{\#blacklivesmatter} and \textbf{\#alllivesmatter} are both liberal-leaning. In~\cite{gallagher2018divergent},
it was pointed out that \textbf{\#alllivesmatter} was used as a counterprotest hashtag between August 2014 and August 2015. This observation helps illustrate that the polarities of tokens can change with time. Nowadays, \textbf{\#bluelivesmatter} is used more than \textbf{\#alllivesmatter} as an antonym of \textbf{\#blacklivesmatter} in practice (in the sense of having a similar semantic meaning but opposite political polarity). Additionally, \textbf{\#alllivesmatter} now appears 
commonly in topics such as animal rights.


\subsubsection{Semantic Components}

To demonstrate the quality of the semantic components $\mathbf{z}^{(s)}$, we calculate the cosine similarity of the embedding vectors of the tokens. Our results appear to be reasonable. For example, we observe that the closest token to ``\textbf{gun}'' is ``\textbf{firearm}'' and that the closest token to ``\textbf{healthcare}'' is ``\textbf{care}''. The t-SNE values from our \textbf{Polarized} \textbf{\modelname} model and \textbf{Complete} \textbf{\modelname} model also suggest that these semantic components have reasonable quality.

\begin{figure}
  \begin{subfigure}[b]{0.45\columnwidth}
    \includegraphics[width=\linewidth]{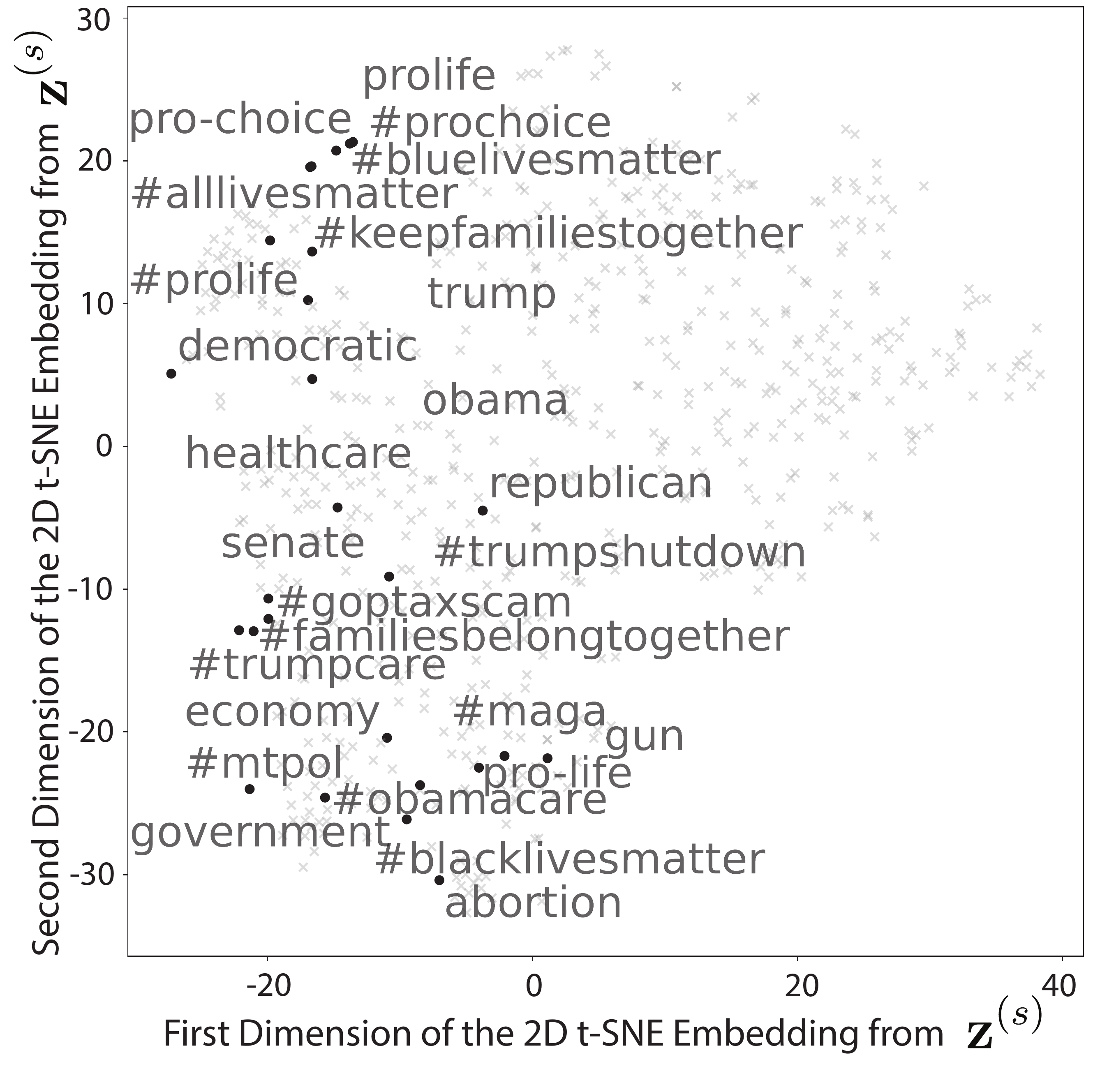}
    \caption{\textbf{Complete} \textbf{\modelname} semantic components.}
    \label{fig:vis_semantic_dim}
  \end{subfigure}
  \hfill 
  \begin{subfigure}[b]{0.45\columnwidth}
    \includegraphics[width=\linewidth]{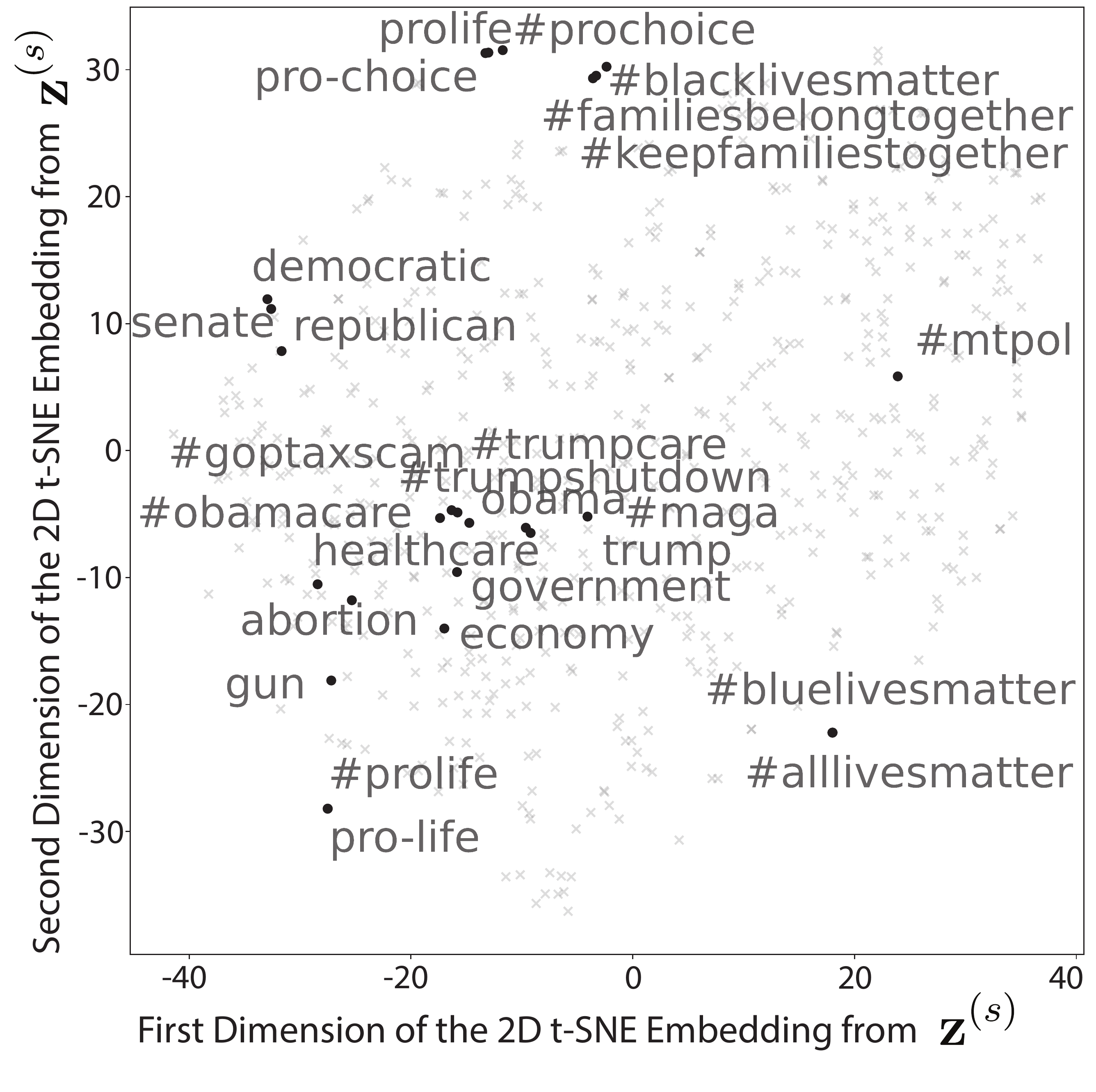}
    \caption{\small \textbf{Polarized} \textbf{\modelname} semantic components.}
    \label{fig:vis_semantic_dim_polarized}
  \end{subfigure}
  \caption{\small Visualization of the semantic components of our (a) \textbf{Complete} and (b) \textbf{Polarized} \textbf{\modelname} embeddings. We project these components onto a plane by calculating t-SNE values. Both results are reasonable, but the \textbf{Polarized} \textbf{\modelname} results tend to encourage semantically-related words to be closer to each other. For example, \textbf{\#familiesbelongtogether} and \textbf{\#keepfamiliestogether} are used similarly in practice and they are close to each other in the embedding from our \textbf{Polarized} \textbf{\modelname} model.}
\end{figure}

In Figure~\ref{fig:vis_semantic_dim}, we plot the results of calculating t-SNE values to project the semantic dimensions of the most-frequent $600$ tokens and several manually-selected tokens from our \textbf{Complete {\modelname}} embeddings onto a plane. In Figure~\ref{fig:vis_semantic_dim_polarized}, we show the t-SNE values for our \textbf{Polarized {\modelname}} embeddings.
These plots illustrate similarities in the semantic meanings of these tokens. For example, we observe that \textbf{\#AllLivesMatter} and \textbf{\#BlueLivesMatter} have similar meanings.
By comparing Figures \ref{fig:vis_semantic_dim} and \ref{fig:vis_semantic_dim_polarized}, it seems that the semantic components of our \textbf{Polarized {\modelname}} embeddings may be slightly more reasonable than those of our \textbf{Complete {\modelname}} embeddings.


\subsubsection{Account-Level Case Studies}\label{subsubsec::acctlvl_cst}

\begin{figure}[h]
\centering
\includegraphics[width=1.0\columnwidth]{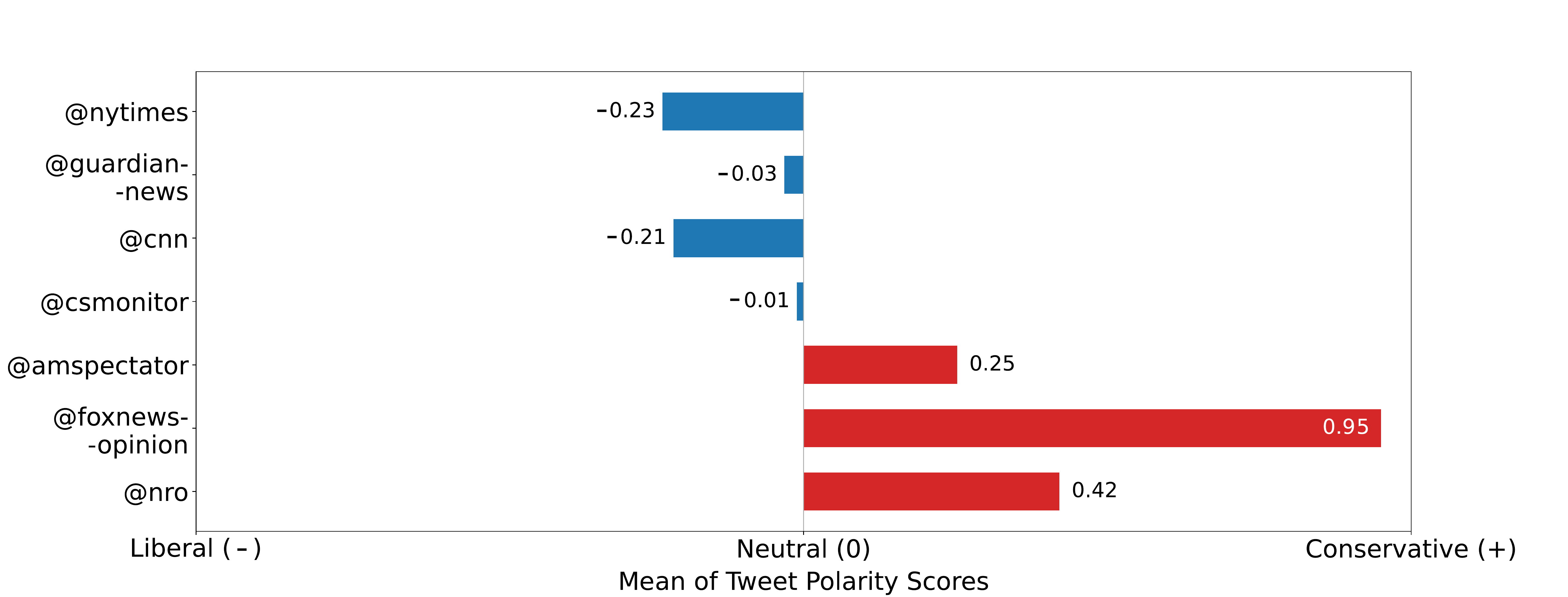} 
\caption{Our estimates of the political polarities of news outlets based on their most recent 3,200 tweets. We collected these tweets starting on 3 March 2019.
}
\label{fig:all_tweets_avg}
\end{figure}

We compute a Twitter account's political polarity by calculating the mean of the polarity scores of all of its tweets. Suppose that an account posted $N$ tweets. The $i^\mathrm{th}$ tweet consists of $n$ tokens, with embeddings $\{\mathbf{z}_1, \ldots, \mathbf{z}_n\}$ and polarity scores $\{\mathbf{z}_1^{(p)}, \ldots, \mathbf{z}_n^{(p)}\}$. The tweet-level polarity score of this tweet is $b_i = (\sum_{j=1}^n \mathbf{z}_j^{(p)}) / n$. We estimate the overall polarity score of the account to be $b = (\sum_{i=1}^{N} b_i)/{N}$. If $b_i < 0$, we regard account $i$ as liberal-leaning; if $b_i > 0$, we regard it as conservative-leaning; if $b_i=0$, we regard it as neutral. We show our results (which seem reasonable) in Figure~\ref{fig:all_tweets_avg}. We plot liberal-leaning accounts in blue and conservative-leaning accounts in red. 

Some previous research~\cite{gu2016ideology,tien2020online} on relationships (e.g., following and retweeting relationships) between Twitter accounts has inferred clearer polarities in news outlets than what we obtain using our approach. This suggests that interactions may be more helpful than text itself at identifying the political polarities of Twitter accounts.


\subsubsection{Illustrations of Estimating Tweet Polarities with the Attention Mechanism}

\begin{figure}[h]
\centering
\includegraphics[width=0.97\columnwidth]{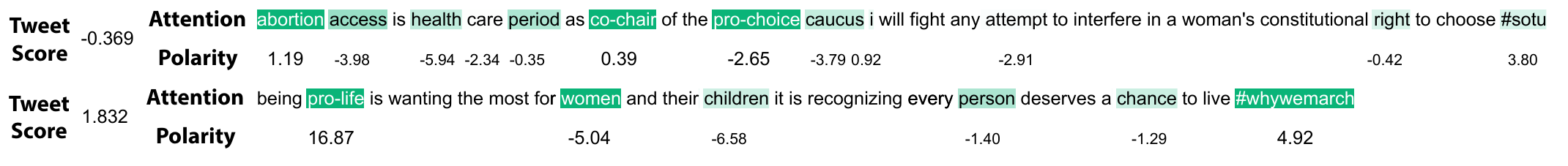}
\caption{\small Illustrations of estimating tweet polarities using an attention mechanism. We show the weights from our \textbf{Complete} \textbf{\modelname} model in green, where darker shades signify greater importance levels. We show the polarity scores underneath the entities and hashtags.
}
\label{fig:attention_weight_visualization}
\end{figure}

See Figure~\ref{fig:attention_weight_visualization} for examples of our \textbf{Complete {\modelname}} model's attention weights and polarity scores. Both the attention weights and the polarity scores appear to be reasonable.


\subsubsection{An Ablation Study of the Attention Mechanism} %

We summarize the performance of the three versions of our \textbf{\modelname} model in Table~\ref{tab:performance}. The left column gives our classification results when we use an attention mechanism. Recall that our \textbf{Baseline} model does not use an attention mechanism. In models with an attention mechanism, we use the score that we infer from Task \#2, which calculates a weighted average of the tokens' political-polarity component $\{\mathbf{z}^{(p)}\}$. In the right column, we show the accuracy and $F_1$ scores when we use the mean value of the elements of $\{\mathbf{z}^{(p)}\}$. Recall that we interpret tweets with negative scores as liberal and tweets with positive scores as conservative.

The results in Table~\ref{tab:performance} suggest that Task \#2 alone can successfully capture polarity information, but introducing Task \#3 to enhance the independence of the semantic and polarity components can improve inference of the political-polarity component $\mathbf{z}^{(p)}$. However, forcing $\mathbf{z}^{(s)}$ to be polarity-neutral makes it harder to preserve accurate semantic information. (See Figures~\ref{fig:vis_semantic_dim} and ~\ref{fig:vis_semantic_dim_polarized}.) This illustrates why our \textbf{Complete} \textbf{\modelname} model does not always outperform our \textbf{Polarized} \textbf{\modelname} model.


\subsection{Results on a Few Downstream Tasks}\label{sub:explore_embeddings_usages}

We illustrate that our embeddings are reliable and useful for several downstream tasks.


\subsubsection{Classification Results}\label{subsubsec:unobserved_accounts_classify}

First, we discuss the classification results of our \textbf{Polarized} and \textbf{Complete {\modelname}} models.

\begin{table} [h]
\centering
\caption{
The classification performance on the withheld tweets of politicians and on the Twitter accounts of politicians. The subscript ``$\mathrm{no\,attn}$'' signifies that we use the mean value of $\{\mathbf{z}^{(p)}\}$ directly (i.e., without applying an attention mechanism). {\sc Skip-Gram} (i.e., the \textbf{Baseline {\modelname}} model) and {\sc GloVe} use a pretrained embedding with the same MLP binary classifier as in our discriminator. (To train this classifier, we use a training set that includes 80\% of the politicians' tweets.) In each entry, we show the accuracy followed by the $F_1$ score. 
We show the best results for each column in bold.
The names of our models are also in bold.
}
{\scriptsize
\begin{tabular}{c|c|c}
\toprule
  \textbf{Model} & \textbf{Tweet-Level Results (accuracy; $F_1$)} & \textbf{Account-Level Results (accuracy; $F_1$)} \\
\midrule
    {\sc Skip-Gram} 
    & 0.7705; 0.7736 & 0.8769; 0.8797 \\
    {\sc GloVe} & 0.7438; 0.7453 & 0.8578; 0.8620 \\
    {\sc BERT$_{\mathrm{base}}$} & \textbf{0.8595}; \textbf{0.8603} & \textbf{0.9965}; \textbf{0.9968}  \\ 
    {\sc BERTweet} & 0.8399; 0.8435 & 0.9844; 0.9853  \\ 
\midrule
    \textbf{Polarized {\modelname}}$_{ \mathrm{no\,attn} }$ & 0.7681; 0.7682 & 0.9757; 0.9758 \\
    \textbf{Complete {\modelname}}$_{ \mathrm{no\,attn}  }$ & 0.7991; 0.7994 & 0.9827; 0.9827 \\
    \midrule
    \textbf{Polarized} \textbf{\modelname} & 0.8339; 0.8337 &  0.9861; 0.9872 \\
    \textbf{Complete} \textbf{\modelname} & 0.8338; 0.8330 & 0.9931; 0.9936 \\
\bottomrule

\end{tabular}
}
\label{tab:performance}
\end{table}

\begin{table} [h]
\centering
\caption{
The classification performance on the unobserved accounts. We never include tweets from these accounts in a training data set. In each entry, we show the accuracy followed by the $F_1$ score. We show the best results for each column in bold. The names of our models are also in bold.
}
{\scriptsize
\begin{tabular}{c|c|c}
\toprule
    \textbf{Model} & \textbf{Tweet-Level Results (accuracy; $F_1$)} & \textbf{Account-Level Results (accuracy; $F_1$)} \\ \midrule
    {\sc Skip-Gram} & 0.5822; 0.5636 & 0.6660; 0.6604 \\
    {\sc GloVe} & 0.5680; 0.5491 & 0.6486; 0.6372 \\ 
    {\sc BERT$_{\mathrm{base}}$} & \textbf{0.6541}; 0.6280 & 0.7234; 0.7218 \\ 
    {\sc BERTweet} & 0.6284; 0.6486 & 0.7836; 0.7778  \\ 
    \midrule
    \textbf{Polarized {\modelname}}$_{ \mathrm{no\,attn} }$ & 0.6066; 0.6244 & 0.8157; 0.8196 \\
    \textbf{Complete {\modelname}}$_{ \mathrm{no\,attn} }$ & 0.6061; 0.6258 & 0.8494; 0.8475 \\
    \midrule
    \textbf{Polarized {\modelname}} & 0.6308; 0.6965 & 0.8493; 0.8758 \\
    \textbf{Complete {\modelname}} & 0.6479; \textbf{0.6987} & \textbf{0.8612}; \textbf{0.8870} \\
\bottomrule
\end{tabular}
}
\label{tab:performance_classifiers}
\end{table}

We select 10\% of the politicians' tweets (there are 127,143 such tweets) uniformly at random and withhold these tweets as the testing set for Table~\ref{tab:performance}. We select another 10\% of the tweets, which we also choose uniformly at random, as a validation set. We use the remaining 80\% of the tweets (i.e., 1,017,137 tweets) as our training set. We train all models (see Table~\ref{tab:performance} and Table~\ref{tab:performance_classifiers}) on the same training set.

In Table~\ref{tab:performance}, we show the performance of the models on the testing set. We perform the tweet-level classification task on the withheld tweets of the politicians. We never include these tweets in the training set. We perform the account-level classification task on the accounts of all politicians with tweets in the testing set. For a given account, we use its tweets in the testing set to infer its political score by calculating the mean polarity score of all of its tweets.

In Table~\ref{tab:performance_classifiers}, we show the tweet-level and account-level classification performance levels for the unobserved accounts. (See Section~\ref{subsec::dataset} for a description of these accounts.)

We use the {\sc Skip-Gram} and {\sc GloVe} embeddings as baselines. For each of these embeddings (which we do not adjust), we use the same MLP classifier that we use as a discriminator in Task \#3 and train the MLP classifiers on our training set until they converge. We fine-tune the transformer classifiers {\sc BERT$_\mathrm{base}$}~\cite{devlin2018bert} and {\sc BERTweet}~\cite{nguyen-etal-2020-bertweet} (which uses the BERT$_\mathrm{base}$ model configuration and is trained using {\sc RoBERTa}-style pretraining) on our training set as baselines. We use the uncased (i.e., ignoring capitalization) version of {\sc BERT$_\mathrm{base}$}; the classifier {\sc BERTweet} separates lower-case and upper-case letters. We use the fine-tuned transformers to classify the tweets of politicians (see Table~\ref{tab:performance}) and the tweets of the unobserved accounts (see Table~\ref{tab:performance_classifiers}).

For the model variants that do not incorporate attention, we calculate each polarity score by computing the mean values of the polarity components $\mathbf{z}^{(p)}$ of the entities and hashtags. We compute the polarities of accounts in the same way as in our examples with news outlets (see Section \ref{subsubsec::acctlvl_cst}).

By comparing Table~\ref{tab:performance} and Table~\ref{tab:performance_classifiers}, we conclude that our models perform better than the transformers ({\sc BERT$_\mathrm{base}$} and {\sc BERTweet}) on the unobserved accounts.
Possible reasons include the following:
\begin{enumerate}
    \item Our polarity score can take any real value, so it can highlight extremists and exploit extreme tweets that help expose an account's polarity. {\sc BERT$_\mathrm{base}$} only allows polarity values between $0$ and $1$. 
    \item Models, such as the transformers, with many
    parameters can suffer from severe overfitting problems, especially when a training set is too small. In Section \ref{sec::limit}, we discuss potential drawbacks of a training data set that includes tweets only by politicians.
\end{enumerate}


\subsubsection{Classification Results using Only Semantic Components}

To demonstrate that including Task \#3 allows the polarity component $\mathbf{z}^{(p)}$ to capture more political information and makes the semantic components $\mathbf{z}^{(s)}$ more politically neutral, 
we conduct an experiment in which we use only the semantic components of the tokens for a classification task. Specifically, we examine account-level classification of the politicians' withheld tweets (see Table~\ref{tab:semantic_classifier}). 

In the left column of Table~\ref{tab:semantic_classifier}, we show our account-level classification results using only $\mathbf{z}^{(s)}$. We obtain these results by training a discriminator with the same architecture as in \textbf{Task \#3}. We train it on our training set (which has 80\% of the politicians' tweets) until the classifier converges on our validation set (which has 10\% of politicians' tweets). We 
then use it to classify tweets in the testing set (which has 10\% of politicians' tweets).

Of our 
classification tasks in Section~\ref{subsubsec:unobserved_accounts_classify},
doing account-level classification based on the politicians' tweets in the testing set is the least challenging one. 
For more challenging classification tasks, such as the classification of the tweets of the unobserved accounts, the accuracies that we obtain by using {\sc Skip-Gram} (i.e., the \textbf{Baseline {\modelname}} model), the\textbf{Polarized {\modelname}} model, and the \textbf{Complete {\modelname}} model are
0.5701, 0.5809, and 0.5756, respectively. Their accuracies for classifying the unobserved 
accounts are 0.6450, 0.6624, and 0.6551, respectively. These numerical values suggest that their performance levels are similar on these tasks. 

\begin{table} [h]
\centering
\caption{The account-level classification performance on the politicians' withheld tweets in our testing set. We never include these tweets in our training data set, but our training set does include other tweets by the accounts that posted these tweets. In each entry, we show the accuracy followed by the $F_1$ score. We show the best results for each column in bold. The names of our models are also in bold. The {\sc Skip-Gram} row indicates our \textbf{Baseline {\modelname}} results.
}
{\scriptsize
\begin{tabular}{c|c|c}
\toprule
    \textbf{Model} & \textbf{Results Based on $\mathbf{z}^{(s)}$ (accuracy; $F_1$)} & \textbf{Results Based on $\mathbf{z}^{(p)}$ (accuracy; $F_1$)} \\ \midrule
    {\sc Skip-Gram} & 0.8394; 0.8451 & 0.8457; 0.8503 \\
    \midrule
    \textbf{Polarized {\modelname}} & \textbf{0.8994}; \textbf{0.9008} & 0.9861; 0.9872 \\
    \textbf{Complete {\modelname}} & 0.8111; 0.8204 & \textbf{0.9931}; \textbf{0.9936} \\
\bottomrule
\end{tabular}
}
\label{tab:semantic_classifier}
\end{table}

The results in Table~\ref{tab:semantic_classifier} suggest that the design of our \textbf{Complete {\modelname}} model helps encourage political information to be in the polarity component $\mathbf{z}^{(p)}$, rather than in the semantic components $\mathbf{z}^{(s)}$.


\subsubsection{Polarity Distribution of Politicians}

\begin{figure}[h]
\centering
\includegraphics[width=0.95\columnwidth]{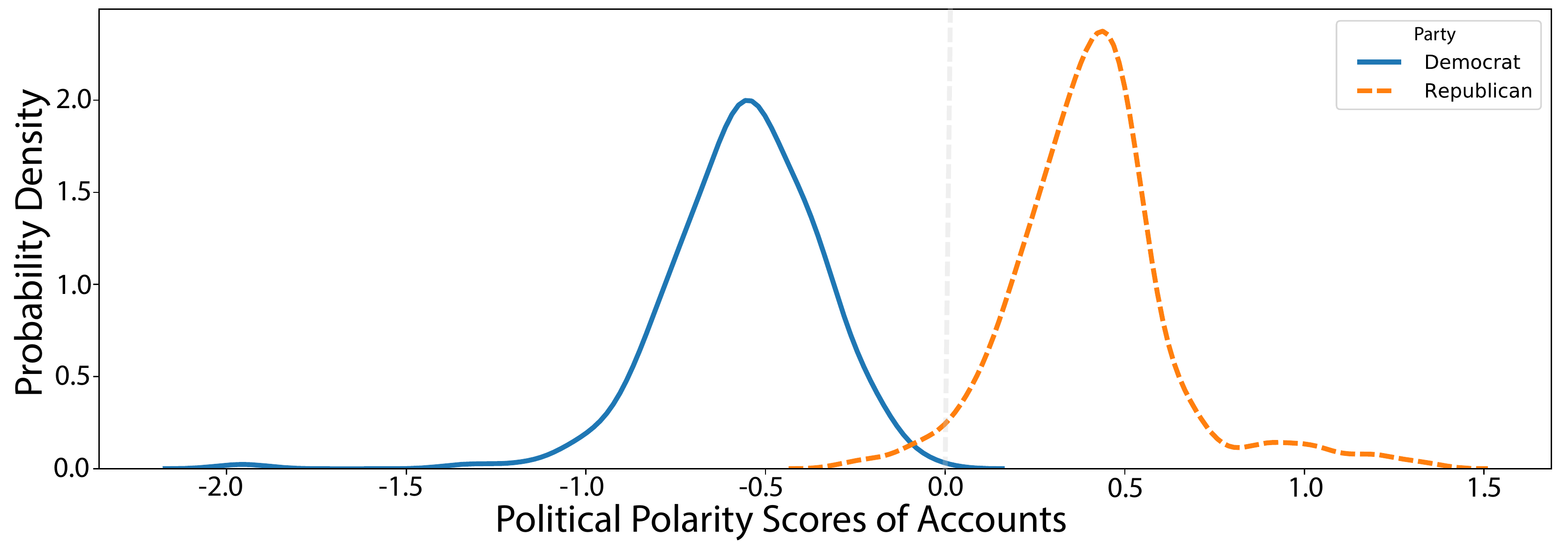}
\caption{Probability densities of the polarity scores of the Twitter accounts of politicians.
}
\label{fig:politicians_distribution}
\end{figure}

We use the same approach as in Section~\ref{subsubsec:unobserved_accounts_classify}
to estimate the polarity scores of the Twitter accounts of politicians. We plot the associated probability densities for both Democrats and Republicans in Figure~\ref{fig:politicians_distribution}, and we observe stark polarization. 


\subsubsection{Temporal Variation of Political Polarities}

We now examine temporal changes in the inferred political polarities of the 49,428 Twitter accounts in the TIMME data set~\cite{10.1145/3394486.3403275} that tweeted in 2020. To examine such temporal variation, we chunk the tweets from 2020 of each of these accounts in 7-day intervals starting from 1 January and examine trends over time. (The final interval is cut off and is hence shorter.)

\begin{figure}[h]
\includegraphics[width=1.0\linewidth]{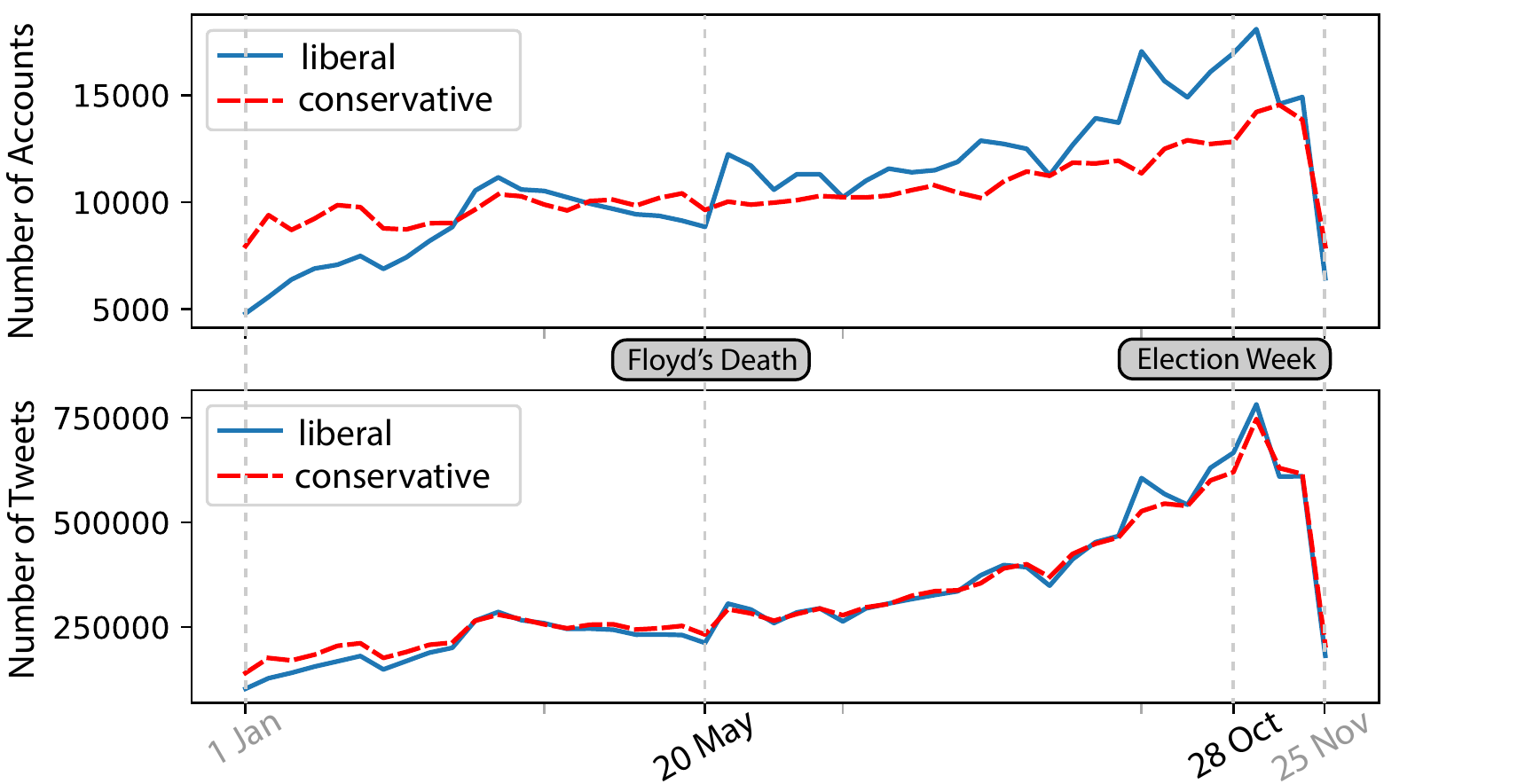}
\vspace{-0.5cm}
\caption{Weekly trends of liberal and conservative tweets on Twitter in 2020. We plot these trends at both (top) the account level and (bottom) the tweet level. The week of George Floyd's murder began on 20 May 2020. The week of the 2020 United States presidential election began on 28 October.
}
\label{fig:time_variance_2020}
\vspace{-0.3cm}
\end{figure}

We use the same approach as in Section~\ref{subsubsec:unobserved_accounts_classify} to infer
tweet-level and account-level polarities.
As we can see in Figure~\ref{fig:time_variance_2020}, our embedding results illustrate plausible trends on Twitter. Many liberal-leaning accounts were active starting in the week of the murder of George Floyd. As the week of the U.S. presidential election approached, people were using Twitter more actively, and then discussions of the election seemed to recede after it was over.
Based on our results, we also suspect that there may be more liberal-leaning accounts than conservative-leaning accounts on Twitter.


\subsubsection{Geographic Distribution of Political Polarities}

\begin{figure}[h]
\centering
\includegraphics[width=\linewidth]{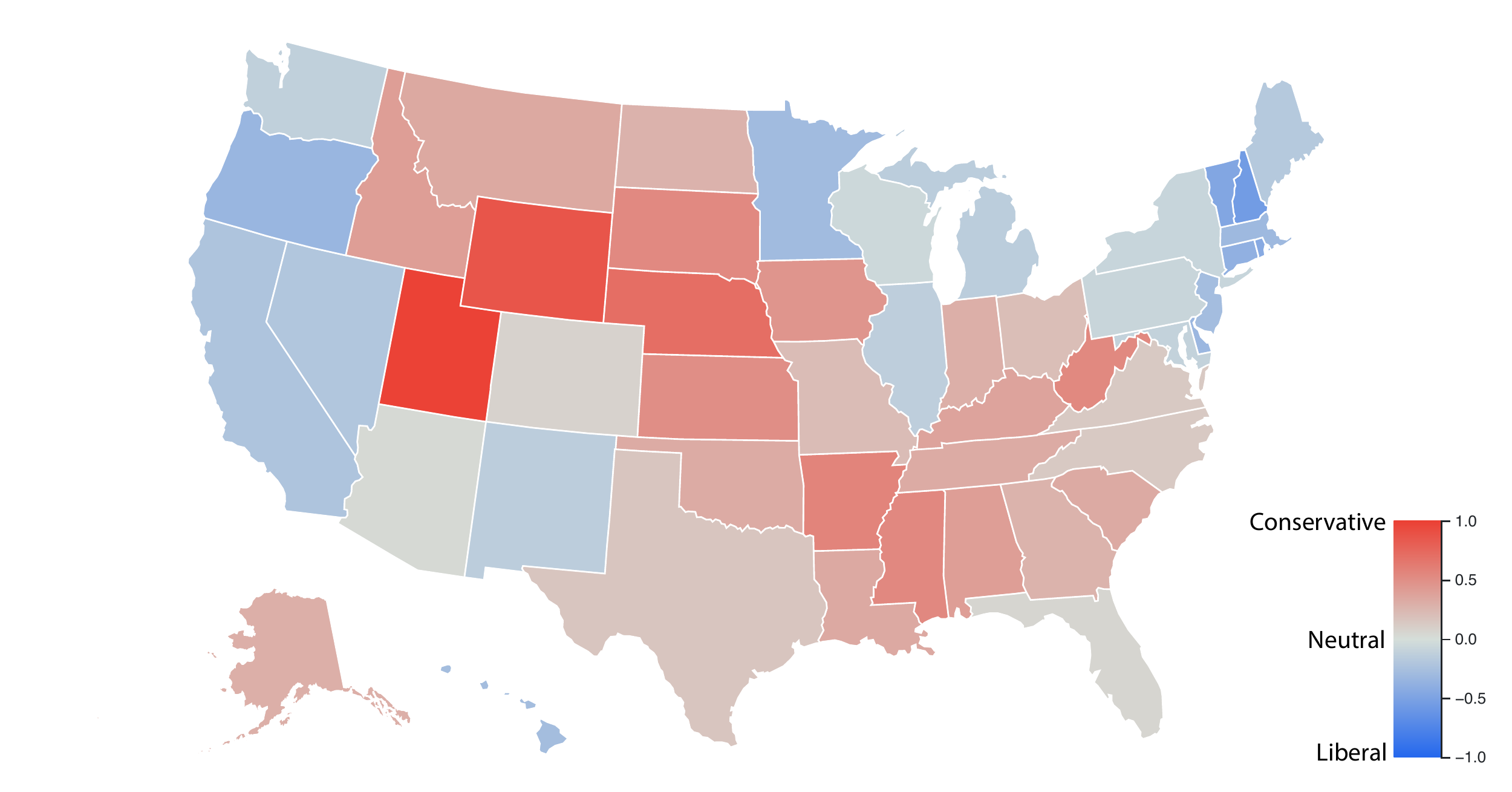}
\vspace{-.5cm}
\caption{The mean polarity score of the Twitter accounts in each state (and the geographic regions Washington, D.C., Puerto Rico, and Guam) in the United States. We normalize the polarity scores to $[-1, 1]$.
}
\label{fig:vis_map_mean_polarity}
\end{figure}

The TIMME data set \cite{10.1145/3394486.3403275} has 51,060 accounts with self-reported geographic locations in the United States. Using these locations, we examine the liberal versus conservative tendencies of tweets across the U.S. in 2020. We calculate the polarity of each Twitter account using the mean of the polarities of the tokens in its tweets; we show these 
account polarities geographically in Figure~\ref{fig:vis_map_mean_polarity}. We use the mean polarity of all accounts in a state (and the geographic regions Washington, D.C., Puerto Rico, and Guam) to calculate the state's polarity, and we then normalize the states' polarity scores $\mathbf{q} = \{q_1, \ldots, q_{53} \}$ to the interval $[-1, 1]$ by calculating $\hat{q}_i = (q_i - \frac{\sum_{j=1}^{53} q_j}{53}) / \max\{|q_1|, \ldots, |q_{53}|\}$. After this normalization, $-1$ is the most liberal score and $+1$ is the most conservative score. Our results are consistent with the tendencies that were reported in national polls for the 2020 U.S. election.\footnote{See \url{https://www.realclearpolitics.com/epolls/2020/president/National.html}.}

\subsubsection{Revealing Biases in Data Sets}\label{subsubsec::reveal_dataset_bias}

\begin{figure}[h]
\centering
\vspace{-0.2cm}
\includegraphics[width=1.0\columnwidth]{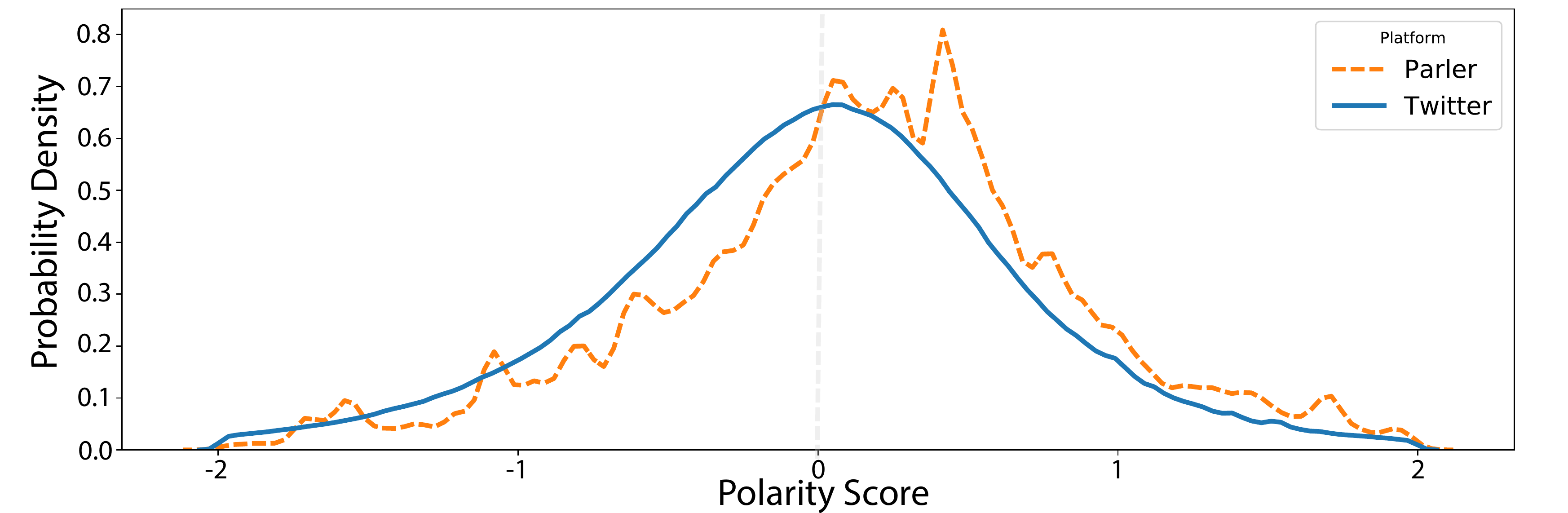}
\caption{Distributions of polarity scores of Twitter tweets and Parler posts. The Twitter curve is smoother because the Twitter data set is much larger than the Parler data set.
}
\label{fig:parler_twitter}
\end{figure}

We use the embedding results of our \textbf{Complete} \textbf{\modelname} model to examine biases in data sets. In practice, using these results entails assuming that we can trust the polarities that we learn from the coarse-grained labels of the politicians' parties. Under this assumption, we find that the TIMME data set is politically neutral and that the {\sc Election2020} data set \cite{chen2020election2020} is somewhat liberal-leaning. In the {\sc Election2020} data set, the mean polarity of the tweets in each week is liberal-leaning. Of the $119$ keywords that were provided in Version 1 of this data set, there are $78$ liberal-leaning keywords and $41$ conservative-leaning keywords. 
Our embedding also suggests that posts on Parler tend to be more conservative than tweets on Twitter. In Figure~\ref{fig:parler_twitter}, we plot the distributions of the polarities of the Twitter tweets and Parler 
posts. We compute these empirical probability densities using kernel density estimation (KDE) with a Gaussian kernel (i.e., the default setting) in the {\sc Seaborn} library~\cite{waskom2021seaborn}.

\subsection{Performance Robustness}

In Table~\ref{tab:performance} and Table~\ref{tab:performance_classifiers}, we reported our best performance levels (from six different random seeds). We also want to examine the robustness of these performance levels. We use the same hyperparameter settings as before, but now we use 5-fold cross validation and different random seeds to initialize the models.

We still train the models on the politicians' tweets. However, instead of randomly using $80$\% of them as our training set, we now do a $5$-fold cross validation. That is, we split the politicians' tweets evenly and uniformly at random into $5$ sets that we select uniformly at random, and we withhold one set at a time as our validation and testing sets (with $10\%$ each, with the tweets in them selected uniformly at random). None of the training sets are identical to the one that we used previously. 

After training a model on the training set, we evaluate it on the testing data set of politicians. We then use the trained models to infer the polarities of the tweets from the unobserved accounts 
using the approaches in
Table~\ref{tab:performance_classifiers}.

\begin{table} [h]
\centering
\caption{The mean values and standard deviations for 5-fold cross validation of different models, which we initialize with different random seeds. We show the best results for each column in bold. The names of our models are also in bold.
}
{\scriptsize
\begin{tabular}{c|c|c}
\toprule
    \multicolumn{3}{c}{Politicians' Accounts (Mean Value $\pm$ Standard Deviation)} \\ \midrule
    \textbf{Model} & \textbf{Tweet-Level Results (accuracy; $F_1$)} & \textbf{Account-Level Results (accuracy; $F_1$)} \\ \midrule
    {\sc Skip-Gram} & $0.7700 \pm 0.0026$ ; $0.7707 \pm 0.0029$  & $0.8833 \pm 0.0113$ ; $0.8996 \pm 0.0100$  \\
    {\sc GloVe} & $0.7231 \pm 0.0039$ ; $0.7319 \pm 0.0035$  & $0.8575 \pm 0.0205$ ; $0.8798 \pm 0.0161$  \\ 
    {\sc BERT$_{\mathrm{base}}$} & $\textbf{0.8586} \pm 0.0006$ ; $\textbf{0.8587} \pm 0.0006$  & $\textbf{0.9963} \pm 0.0034$ ; $\textbf{0.9963} \pm 0.0034$  \\ 
    {\sc BERTweet} & $0.8337 \pm 0.0010$ ; $0.8327 \pm 0.0010$ & $0.9828 \pm 0.0077$  ; $0.9826 \pm 0.0077$  \\ 
    \midrule
    \textbf{Polarized {\modelname}}$_{ \mathrm{no\,attn} }$ & $0.7691 \pm 0.0011$ ; $0.7665 \pm 0.0011$ & $0.9721 \pm 0.0244$ ; $0.9723 \pm 0.0243$ \\
    \textbf{Complete {\modelname}}$_{ \mathrm{no\,attn} }$ & $0.7955 \pm 0.0009$ ; $0.7937 \pm 0.0009$ & $0.9805 \pm 0.0169$ ; $0.9811 \pm 0.0167$ \\
    \midrule
    \textbf{Polarized {\modelname}} & $0.8338 \pm 0.0007$ ; $0.8336 \pm 0.0007$  & $0.9841 \pm 0.0030$ ; $0.9845 \pm 0.0030$ \\
    \textbf{Complete {\modelname}} & $0.8332 \pm 0.0006$ ; $0.8327 \pm 0.0006$  & $0.9915 \pm 0.0026$ ; $0.9927 \pm 0.0026$  \\
\toprule
    \multicolumn{3}{c}{Unobserved Accounts (Mean Value $\pm$ Standard Deviation)} \\ \midrule
    \textbf{Model} & \textbf{Tweet-Level Results (accuracy; $F_1$)} & \textbf{Account-Level Results (accuracy; $F_1$)} \\ \midrule
    {\sc Skip-Gram} & $0.5822 \pm 0.0007$ ; $0.5635 \pm 0.0008$  &  $0.6561 \pm 0.0053$ ; $0.6324 \pm 0.0074$  \\
    {\sc GloVe} & $0.5764 \pm 0.0009$ ; $0.5574 \pm 0.0009$ & $0.6387 \pm 0.0073$ ; $0.6222 \pm 0.0099$  \\ 
    {\sc BERT$_{\mathrm{base}}$} & $0.6348 \pm 0.0007$ ; $0.6231 \pm 0.0006$  & $0.7182 \pm 0.0078$ ; $0.7149 \pm 0.0072$  \\ 
    {\sc BERTweet} & $0.6282 \pm 0.0006$ ; $0.6280 \pm 0.0005$ & $0.7752 \pm 0.0176$ ; $0.7695 \pm 0.0173$  \\ 
    \midrule
    \textbf{Polarized {\modelname}}$_{ \mathrm{no\,attn} }$ & $0.6245 \pm 0.0011$ ; $0.6067 \pm 0.0011$  & $0.8062 \pm 0.0191$ ; $0.8105 \pm 0.0182$  \\
    \textbf{Complete {\modelname}}$_{ \mathrm{no\,attn} }$ & $0.6259 \pm 0.0014$ ; $0.6063 \pm 0.0015$  & $0.8467 \pm 0.0177$ ; $0.8450 \pm 0.0178$  \\
    \midrule
    \textbf{Polarized {\modelname}} & $0.6284 \pm 0.0023$ ; $0.6865 \pm 0.0020$  & $0.8463 \pm 0.0063$ ; $0.8666 \pm 0.0059$ \\
    \textbf{Complete {\modelname}} & $\textbf{0.6472} \pm 0.0030$ ; $\textbf{0.6907} \pm 0.0028$ & $ \textbf{0.8550} \pm 0.0075$ ; $\textbf{0.8814} \pm 0.0072$  \\
\bottomrule
\end{tabular}
}
\label{tab:performance_stability}
\end{table}

In Table~\ref{tab:performance_stability}, we report the means and standard deviations from our 5-fold cross validation.
The results illustrate that the models' performance levels are robust, although the tweet-level performance levels are more robust than the account-level performance levels.


\subsection{Bot Analysis}

Our investigation does not account for the activity of automated accounts (i.e., bots). We use the verified Twitter accounts of politicians, so we assume that these are not bot accounts.
However, 
bots are widespread on Twitter and other social media~\cite{ferrara2016rise},
We check for potential bots in our Twitter accounts and compare the inferred bot probabilities of these accounts with our inferred political polarities.
We find that the probability that an account is a bot has little correlation with its political polarity.

To evaluate the probability that a Twitter account is a bot, we use Botometer (version 4)~\cite{sayyadiharikandeh2020detection}. It has two options --- universal and English --- for the language that it employs for bot detection. The universal bot score is evaluated in a language-independent way, but the English bot score is more accurate for accounts that tweet primarily in English, so we use the English option.

\begin{figure}[h]
\centering
\includegraphics[width=0.9\columnwidth]{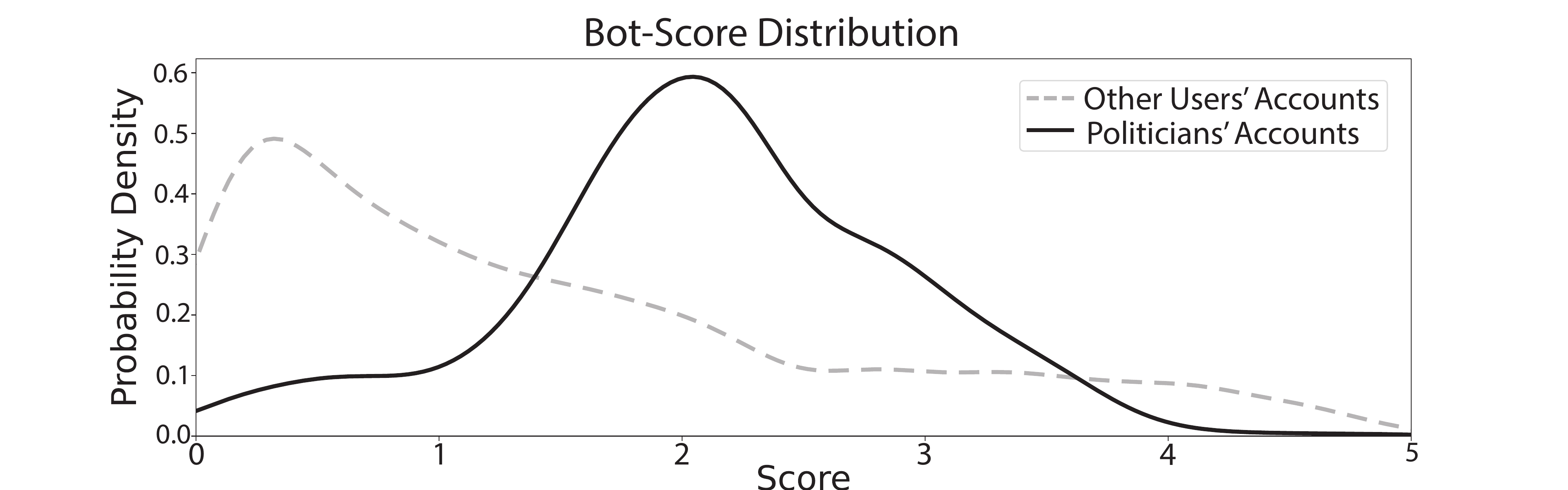}
\caption{Probability densities of the bot scores of the Twitter accounts of politicians (solid curve) and all other Twitter accounts (dashed curve).
}
\label{fig:overall}
\end{figure}

There are many different types of Twitter bots (see \url{https://botometer.osome.iu.edu/faq}). For simplicity, we use only an overall bot score from Botomer. The score of a bot varies between $0$ and $5$, with larger scores signifying that an account is more likely to be a bot. In Figure~{\ref{fig:overall}}, we show the probability densities
of bot scores for politicians and ordinary Twitter accounts.

\begin{figure}[h]
\centering
  \begin{subfigure}[b]{0.45\columnwidth}
    \includegraphics[width=\linewidth]{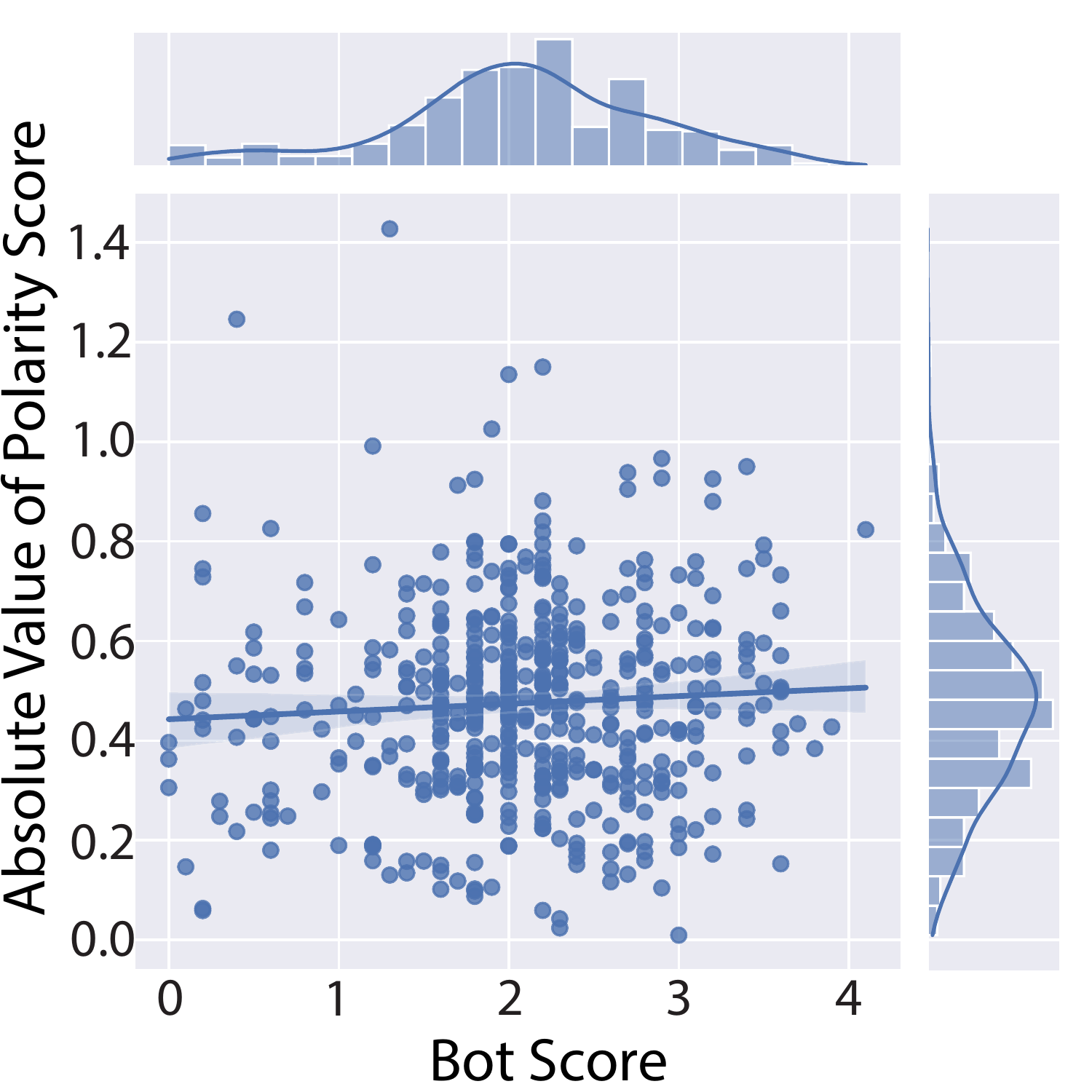}
    \caption{}
    \label{fig:vis_corr_politician}
  \end{subfigure}
  \hfill 
  \begin{subfigure}[b]{0.45\columnwidth}
    \includegraphics[width=\linewidth]{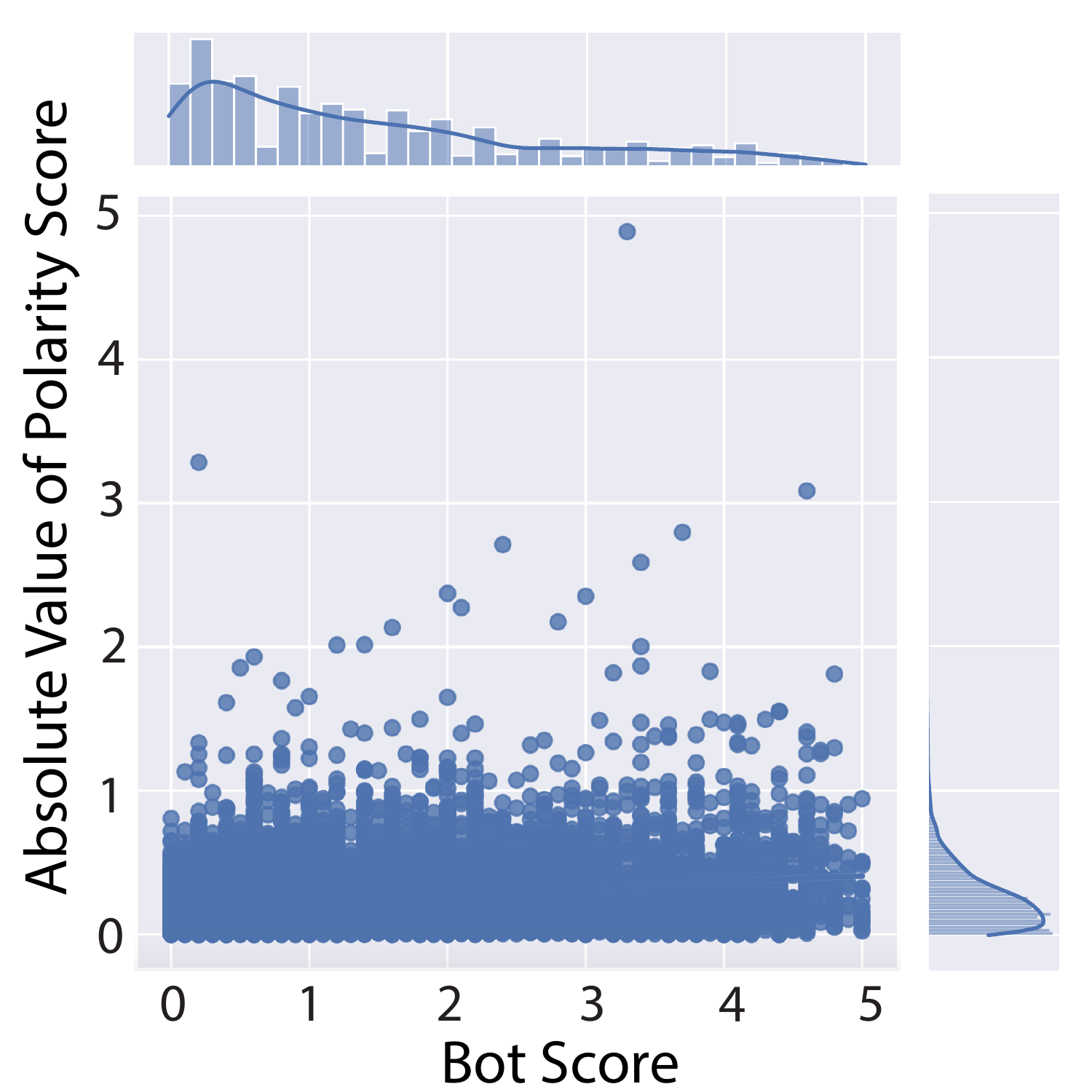}
    \caption{}
    \label{fig:vis_corr_other}
  \end{subfigure}
\caption{Distribution of the overall bot score versus the absolute values of the polarity scores of the content of (a) politicians' Twitter accounts and (b) all other Twitter accounts.
}
\label{fig:distribution_politicians}
\end{figure}

In Figure~\ref{fig:distribution_politicians}, we plot the distributions of the overall bot scores versus the absolute values of polarity scores (i.e., $|\{\mathbf{z}^{(p)}\}|$) for both politicians' Twitter accounts and ordinary Twitter accounts. 
The absolute values of the polarity scores indicate the extremeness of an account's content according to our {\modelname} model.

\subsection{Impact of Assigning Polarity Scores to Other Tokens}


We use tokens other than hashtags and entities in our \textbf{\modelname} model, but we have not assigned political polarities to them. 
We feel that this design decision improves the interpretability of our model. For some words, such as \textbf{``a''} or \textbf{``the''}, it definitely does not make sense to assign a political polarity.

\begin{table} [h]
\centering
\caption{
The tweet-level classification performance on the politicians' withheld tweets in our testing set when we assign polarity scores to all tokens versus only assigning polarity scores to
hashtags and entities. In each entry, we show the accuracy followed by the $F_1$ score. We show the best results for each column in bold. 
}
{\scriptsize
\begin{tabular}{c|c|c}
\toprule
    \textbf{Results (accuracy; $F_1$)} & {Polarized {\modelname}} & {Complete {\modelname}} \\
    \midrule
    {Using $\mathbf{z}^{(p)}$ of All Tokens} & \textbf{0.8369}; \textbf{0.8366} & 0.8337; \textbf{0.8334}  \\
    {Using $\mathbf{z}^{(p)}$ of Only Entities and Hashtags} & 0.8339; 0.8337 & \textbf{0.8338}; 0.8330 \\
\bottomrule
\end{tabular}
}
\label{tab:classifier_without_entities}
\end{table}

As one can see in Table~\ref{tab:classifier_without_entities}, assigning political polarities to tokens other than named entities and hashtags does not seem to harm our classification performance.
We show it by comparing the tweet-level classification results of our \textbf{Complete {\modelname}} model on the withheld testing set of the politicians' tweets (i.e., the same testing set that we used in Section~\ref{subsubsec:unobserved_accounts_classify}). 

%% file: 6-conclusion.tex
\section{Limitations}\label{sec::limit}

We highlight several important limitations of our work. Naturally, our discussion is not exhaustive, and it is also relevant to think about other limitations.


\subsection{Incomplete Data} 

We consider only textual information. Therefore, we overlook images, videos, and other types of information. 


\subsection{Model Limitations} 

We designed our \textbf{\modelname} model to infer political polarity scores from entities and hashtags, so it is not helpful for inferring the polarity of tweets that have no entities or hashtags. Additionally, our \textbf{\modelname} model does not take time stamps into account, so it does not consider the dynamic nature of polarities.


\subsection{Training-Set Biases and Other Issues} \label{subsubsec::bias_and_bots}

Our design decision of assigning political polarities to items in a training set enables one to automatically assign labels at scale. However, it can be undesirable to make such assignments a priori.

We use the tweets of politicians because their accounts are verified and they have a consistent, unambiguous, and self-identified political affiliation. However, this choice introduces biases and other potential issues. First, the size of our training data set is necessarily limited, and it is easier for models to overfit data when using small data sets than when using large ones. Second, our results may be sensitive to the time window in which we collected tweets. For example, polarization in tweets may be more apparent during elections than at other times. Third, politicians are not necessarily representative of other social-media users.
Fourth, we did not train our model to handle bot or cyborg accounts. We used verified Twitter accounts in our training data set, so it presumably does not have any bots or cyborgs. (Our estimation of bot probabilities supports this presumption.) Bot accounts are very common on Twitter~\cite{ferrara2016rise}, so it is necessary to be cautious when applying our model directly to typical Twitter data sets.

The verified Twitter accounts of politicians are very different in nature from the Twitter accounts of other users. We saw ramifications of such differences in our classification results. Using {\sc BERT$_{\mathrm{base}}$} to classify tweets from politicians versus those of other accounts yields an accuracy of 0.7590 and an $F_1$ score of 0.7595 on the testing set. If we partition the set of non-politician accounts into two groups that each have the tweets of 1,293 accounts (which we assign uniformly at random) and try to classify the group of each tweet, we obtain an accuracy of $0.4600$ and an $F_1$ score of $0.6276$.


\subsection{Quantifying Political Polarity} 

There are many possible ways to quantify political polarity. We chose to assign labels of ``liberal'' and ``conservative'', but other dichotomies are also relevant. Moreover, we designed our \textbf{\modelname} model learn a single type of polarity. It cannot simultaneously reveal multiple types of political polarities.


\subsection{Sarcasm and Irony} 

In our work, we did not analyze nuanced situations, such as sarcasm and irony, that depend heavily on context. Sarcasm plays an important role in social media~\cite{tayal2014polarity}, and it is worth generalizing our \textbf{\modelname} model to be able to handle it successfully in the future.


\section{Conclusions}\label{sec::con}

We studied the problem of inferring political polarities in embeddings of entities and hashtags. To capture political-polarity information without using auxiliary word pairs, we proposed \textbf{\modelname}, a multi-task learning model that employs an adversarial framework. 

Our experiments illustrated the effectiveness of our \textbf{\modelname} model and the usefulness of the embeddings that one can produce from it.
In principle, it is possible to extend our approach to extract any type of polarity of an embedding (while attempting to minimize the effects of polarity on other components). One can also extend our \textbf{\modelname} model to deploy it with a variety of embedding strategies.


\section{Ethics Statement}

There are several ethical points to consider in our work.

First, one needs to consider our data sets. The data that we used comes from publicly available sources, and our training data comes from the verified accounts of politicians. We do not store any sensitive information (such as real-time locations) from Twitter. It is important to be aware of Twitter's privacy policy (see \url{https://twitter.com/en/privacy}) when downloading and using data from Twitter.

There are also important ethical considerations when using the results of embeddings like ours. Our \textbf{\modelname} model yields interesting and occasionally counterintuitive results. One must be cautious when using such results for subsequent tasks (e.g., when drawing conclusions about an individual's political views). Additionally, models inherit biases from training data sets, and they can exacerbate such biases~\cite{o2016weapons}.

The conclusions that we obtained from applying our \textbf{\modelname} model are based on the existing posts of social-media accounts. One must be cautious when subsequently inferring what such accounts may post in the future and especially if one seeks to use any insights from our model to inform behavior, actions, or policy.

%% file: 7-acknowledgement.tex

\section*{Declarations}


\subsection*{List of Abbreviations}


\begin{table}[h]
    \caption{List of Abbreviations}
    \centering
    \begin{tabular}{c|c}
    \toprule
        \textbf{Abbreviation} & \textbf{Corresponding Term} \\  \midrule
        PEM & Polarity-aware Embedding Multi-task learning \\
        t-SNE & t-distributed stochastic neighbor embedding  \\
        BERT & Bidirectional Encoder Representations from Transformers  \\
        GloVe & Global Vectors for Word Representation \\
        TIMME & Twitter Ideology-detection via Multi-task Multi-relational Embedding  \\
        NCE & noise-contrastive estimation \\
        KDE & kernel density estimation \\
    \bottomrule 
    \end{tabular}
    \label{tab:list_abbr}
\end{table}

In Table~\ref{tab:list_abbr}, we list the most important abbreviations in our paper.


\subsection*{Ethics Approval and Consent to Participate}

Not applicable.


\subsection*{Consent for Publication}

Not applicable.


\subsection*{Availability of Data and Materials}



Our code, the data sets of the politicians, and the embedding results of our models are available at \url{https://bitbucket.org/PatriciaXiao/pem/src/master/}.


\subsection*{Competing interests}


Not applicable.


\subsection*{Funding}




This research was supported by the National Science Foundation (through grants III-1705169, NSF 1937599, NSF 2119643, and 1922952), an Okawa Foundation Grant, Amazon Research Awards, Cisco research grant USA000EP280889, Picsart Gifts, and Snapchat Gifts.


\subsection*{Authors' Contributions}


ZX, PZ, MAP, and YS conceived and conceptualized the study. 
ZX, JZ, YW, and WHL performed the analysis and wrote the initial draft of the paper.
ZX, MAP, and YS reviewed and extensively edited the manuscript, determined what additional analysis was necessary, and produced the final version of the manuscript.
All authors read and approved the final manuscript.


\subsection*{Acknowledgements}


We thank Yupeng Gu and Zhicheng Ren for helpful discussions.